\documentclass[11pt]{article}
\pdfoutput=1
\usepackage{jcapmod}
\usepackage{mathtools}
\usepackage{microtype}
\usepackage[utf8]{inputenc}
\usepackage[english]{babel}
\usepackage[dvipsnames]{xcolor}
\usepackage{amsmath,amssymb,amsbsy,amstext,amsthm}
\usepackage[colorlinks=true]{hyperref}
\usepackage{microtype}
\usepackage{graphicx}
\usepackage{amsfonts}
\usepackage{upgreek}
\usepackage{exscale,relsize}
\usepackage[makeroom]{cancel}
\usepackage{soul}
\usepackage{float}
\usepackage{empheq}
\usepackage[most]{tcolorbox}
\RequirePackage{color}
\usepackage{feynmp}
\usepackage{ifpdf}
\ifpdf
\fi

\usepackage{colortbl}
\definecolor{green2}{cmyk}{0, 1, 0.5, 0}
\definecolor{lightgreen}{cmyk}{0.2, 0, 0.2, 0.2}
\definecolor{dred}{rgb}{0.9,0.2,0.5}
\definecolor{dred2}{cmyk}{0.1,0.7,0.1,0.3}
\definecolor{lightgray2}{cmyk}{0.4,0.4,0,0.8}
\definecolor{black}{cmyk}{1.0,1.0,1.0,1.0}

\definecolor{verde}{rgb}{0,0.5,0}

\AtBeginDocument{
 \hypersetup{
 urlcolor=NavyBlue,
 citecolor=NavyBlue,
 linkcolor=BrickRed,
 }
}

\allowdisplaybreaks[1]

\usepackage{colortbl}

\makeatletter
\newlength{\apb@width}
\newcommand{\autoparbox}[2][c]{\settowidth{\apb@width}{#2}\parbox[#1]{\apb@width}{#2}}

\makeatother

\numberwithin{equation}{section}

\def\beq{\begin{equation}}
\def\eeq{\end{equation}}

\def\bea{\begin{eqnarray}}
\def\eea{\end{eqnarray}}
\def\eg{{\it e.g.~}}

\def\ie{{\it i.e.~}}
\def\d{{\rm d}}

\def\d{{\rm d}}

\def\nn{\nonumber}

\def\sgm{\sigma}

\def\Mp{M_{\rm pl}}

\def\fr{\frac}

\def\0{{\boldsymbol 0}}

\def\fr{\frac}

\newcommand{\vbf}[1]{\mathbf{#1}}

\newtcbox{\mymath}[1][]{%
    nobeforeafter, math upper, tcbox raise base,
    enhanced, colframe=gray!30!gray,
    colback=gray!10, boxrule=0.5pt,
    #1}

\usepackage{setspace} 

\begin{document}

\begin{titlepage}

\setcounter{page}{1} \baselineskip=15.5pt \thispagestyle{empty}

\bigskip\

\vspace{1cm}
\begin{center}

{\fontsize{19}{28}\selectfont  
{\bf Scale-dependent chirality as a smoking gun for \\\vspace{0.25cm}
Abelian gauge fields during inflation 
}}
\bigskip
\end{center}

\vspace{0.2cm}
\begin{center}
{\fontsize{13}{30}\selectfont Ogan \"Ozsoy
$^{\ddagger}$,
}
{\fontsize{13}{30}\selectfont Alexandros Papageorgiou
$^{\ddagger}$,
}
{\fontsize{13}{30}\selectfont Matteo Fasiello
$^{\ddagger}$
}
\end{center}
\begin{center}
\textsl{$^\ddagger$ Instituto de Física Téorica UAM-CSIC, c/ Nicolás Cabrera 13-15,
28049, Madrid, Spain\textit{}}
\vskip 8pt
\end{center}

\vspace{1.2cm}
\noindent
\begin{abstract}

Axion-inflation models are a compelling candidate as a mechanism behind the accelerated expansion in the early universe  in light of the possibility to embed them in higher dimensional UV complete theories and the exciting prospect of testing them with next-generation cosmological probes. Adding an Abelian gauge sector to axion-inflation models makes for a rich, interesting, phenomenology spanning from primordial black holes to gravitational waves (GWs). Several recent studies employ an approximate analytic (Gaussian) template to characterize the effect of gauge field production on cosmological perturbations. In this work we go beyond such approximation and numerically study particle production and the ensuing scalar and tensor spectra. We find a significant deviation from results based on log-normally distributed vector field excitations. As an important phenomenological application of the improved method, we study the expected chirality and spectral index of the sourced GW background at scales relevant for current and next-generation GW detectors. One striking feature is that of a scale-dependent chirality.
We derive a consistency relation between these two observables that can serve as an important tool in identifying key signatures of multi-field dynamics in axion inflation.

\end{abstract}
\vspace{0.6cm}
 \end{titlepage}

\tableofcontents 

\section{Introduction}

A period of accelerated expansion in the very early universe can be readily implemented by means of a scalar field equipped with a sufficiently flat potential \cite{Lyth:1998xn}. The most constraining data to date on the physics of inflation comes from observations at the largest scales, those of the CMB. The inflationary mechanism has been spectacularly successful  in explaining how quantum fluctuations, magnified by the expansion, gave rise to the observed CMB temperature anisotropies and planted the seeds for those matter inhomogeneities that evolved into the cosmos we observe today. Remarkably, all the information gathered from cosmological probes still places only rather qualitative bounds on the micro-physics of inflation. Key questions such as those on the nature of the inflationary particle content remain unanswered. 

The unprecedented array of cosmological probes set to become operational in the next two decades holds the potential to address such fundamental questions. Probes of the cosmic microwave background such as CMB-S3 \cite{SimonsObservatory:2018koc} and later LiteBIRD \cite{LiteBIRD:2022cnt} and CMB-S4 \cite{CMB-S4:2016ple,CMB-S4:2022ght} will improve by almost two order of magnitudes existing bounds on primordial gravitational waves (PGW). Astronomical surveys (such as Euclid \cite{Amendola:2016saw}, LSST \cite{LSSTScience:2009jmu}) will close in on the key observable for inflationary (self) interactions, the non-linear parameter $f_{\rm NL}$ \cite{Bartolo:2004if}, and  in particular hit the target value $\sigma_{f_{\rm NL}}\sim 1$, a crucial threshold vis-\`{a}-vis the single \textit{vs} multi-field inflation dichotomy. Probes of the 21cm background promise an $\sigma_{f_{\rm NL}}$ further improved by up two orders of magnitude \cite{Munoz:2015eqa} down the line. 

Perhaps the most exciting opportunities will be afforded by current and upcoming gravitational wave probes at intermediate \cite{EPTA:2023xxk,NANOGrav:2023gor,SKA:2018ckk} and small \cite{LIGOScientific:2016wof,LIGOScientific:2021aug,LISACosmologyWorkingGroup:2022jok,Maggiore:2019uih,Reitze:2019iox,Kawamura:2020pcg} scales. These have opened up an entire new window (spanning multiple decades in frequency!) on the physics of the early universe. Since LIGO has become operational one may no longer assume a blank canvas for the PGW signal at small frequencies. Further important input is due to pulsar timing arrays. In this context one ought mention the possibility that the recently detected signal \cite{NANOGrav:2023gor,NANOGrav:2023hvm,Reardon:2023gzh,Xu:2023wog} in the nHz range may indeed be of primordial origin \cite{Ellis:2020ena,Blasi:2020mfx,Blanco-Pillado:2021ygr,NANOGrav:2023hvm,EPTA:2023xxk,Unal:2023srk,Addazi:2020zcj,Gouttenoire:2023bqy}.  
In the space of a few years the canvas will be populated by the sensitivity curves of several pulsar timing  array experiments and GW interferometers. Undoubtedly then, this is the ideal time to put our most compelling inflationary models to the test, focusing in particular on their the PGW spectrum.

In delivering the predictions associated to the inflationary paradigm, one may opt for different approaches. The top-down perspective seeks acceleration mechanisms that are embedded (or embeddable) in UV complete theories such as string theory. This serves as a powerful model selection criterion \cite{Baumann:2014nda}. The complementary, bottom-up, approach is that familiar from effective field theory: a general Lagrangian is written down, compatible with the (putative) symmetries of the system and organised around increasingly irrelevant operators \cite{Cheung:2007st,Weinberg:2008hq}. Naturally, both approaches have born many fruits over the years. In this work we shall focus on axion-inflation models, characterized by axion-like particles (ALPs) that act as the inflaton field or populate hidden axion sector(s) of the inflationary Lagrangian.

There exist a rich literature on supergravity and string theory realizations of axion inflation \cite{Svrcek:2006yi,Pajer:2013fsa}. These constructions may very well comprise also  gauge sectors coupled to axions via e.g. a Chern-Simons (CS) term.  Although such generic particle content can be realized in string theory, the parameter space corresponding to testable GW signatures is most naturally obtained in the case of Abelian gauge fields \cite{Holland:2020jdh,Bagherian:2022mau,Dimastrogiovanni:2023juq}. This configuration is the subject of the present work. One should add that, already at the phenomenological level, axion inflation models posses compelling properties such as an approximate shift symmetry protecting the inflaton mass from large quantum correction thereby tackling the $\eta$ problem. Coupling axions with gauge fields can make for interesting dynamics in the immediately post-inflationary phase \cite{Adshead:2015pva,Hashiba:2021gmn}.

The most striking (as well as most general) observational features associated with the presence of a CS coupling are to be found in the GW sector. The corresponding GW spectrum may indeed exhibit a peak structure and non-trivial chirality of the signal. The position of the peak(s) depends on initial conditions as well as on the field content. An impressive number of upcoming GW probes are set to scan a wide frequency range (from ${\rm nHz}$ to at least ${\rm kHz}$), making the prospect of detection of primordial GW ever more realistic. The chirality of the signal can be tested at large scales via $\langle BT \rangle$ and $\langle EB \rangle$ correlations \cite{Thorne:2017jft,Campeti:2020xwn} and at small scales by (i) considering cross-correlation between different interferometers\footnote{So that the combined effective geometry of the interferometer is non-planar.} \cite{Smith:2016jqs};  and (ii) by means of the  dipolar anisotropy induced by the motion of the solar system with respect to the cosmic rest frame \cite{Domcke:2019zls}.

As was recently shown in \cite{Garcia-Bellido:2023ser}, in inflationary models that contain CS type axion-Abelian gauge field interactions, GW chirality can exhibit non-trivial frequency dependence. This is not due to the standard transition from the vacuum-dominated part of the signal to the range where the sourced contribution takes the lead, but rather to a more subtle non-linear effect in the 1-loop sourcing of the GW power spectrum.  The blue-tilted part of a peak of the GW spectrum corresponds to a rapidly increasing gauge quanta production whose sourcing of primordial GWs is nearly insensitive (see section \ref{sec4p2}) to the specific polarization. {This is because the sourcing in this case is dominated by gauge modes exhibiting relatively large momenta for which the phase space available for the inverse decay process $\delta A + \delta A \to h_\lambda$ is identical for both GW polarization states $\lambda = \pm$.} For a nearly constant or decreasing rate of gauge quanta production, the GW spectrum will instead receive the main contribution from the domain of the 1-loop phase-space integral that is most dependent on the polarization and will therefore exhibit the largest chirality. 

The main purpose of this work is to explore this subtle effect for the case of spectator axions coupled to Abelian gauge fields and clarify how this effect is present already in the small back-reaction regime\footnote{See also \cite{Bastero-Gil:2022fme} for recent work focusing on the directly coupled axion inflaton - Abelian gauge field system.}. Furthermore, we show how it can have important implications for the predicted chirality within reach of current and next generation GW detectors. We also compute the power spectrum of scalar perturbations\footnote{In our computation we disregard the scalar perturbations of the metric, as typically done in the literature. This is justified in light of e.g. \cite{Durrer:2024ibi} which suggests that the metric perturbations only affect the sourced power spectrum for far smaller values of the particle production parameter than those we consider here.} fully numerically, exploiting the numerical scheme of \cite{Garcia-Bellido:2023ser}. We use the numerical solutions for both tensor and scalar spectra as benchmarks to evaluate the validity of the semi-analytical WKB approximation commonly  employed in analysing these scenarios (see also \cite{Corba:2024tfz} for a recent study on cross-correlations).

In light of our findings, the notion that GW sourced via CS-coupled gauge fields are necessarily (maximally) chiral needs to be revisited. As we will show, in the Abelian case it is quite possible to have a slow build up in the chirality across several orders of magnitude in frequency. At the heart of this feature  is the non-linear nature of GW sourcing via Abelian gauge modes. This mechanism being negligible in the non-Abelian scenario \cite{Adshead:2012kp,Dimastrogiovanni:2012ew,Adshead:2016omu,Dimastrogiovanni:2016fuu,Agrawal:2017awz,Caldwell:2017chz,Thorne:2017jft,Dimastrogiovanni:2018xnn,Fujita:2018vmv,Domcke:2018rvv,Lozanov:2018kpk,Watanabe:2020ctz,Holland:2020jdh,Domcke:2020zez,Iarygina:2023mtj,Dimastrogiovanni:2023oid,Ishiwata:2021yne,Durrer:2024ibi,Dimastrogiovanni:2024xvc}, we identify scale-dependent chirality as smoking-gun signature of CS-coupled Abelian gauge fields during inflation. 

This paper is organized as follows: in section \ref{sec:2} we briefly summarize the model under study. We give a detailed account of the WKB approximation which is the usual method  employed to produce semi-analytic results for the power spectra and explain how the standard procedure can yield inaccurate results for the UV and IR tails of the total power spectra. We end the section by suggesting an alternative and improved approach, still within the WKB approximation scheme. In section \ref{sec3} we tackle the power spectra using three levels of approximation. We use the fully numerical method as a benchmark and confront these results with the ones arising from the standard WKB procedure as well as our improved method introduced in section \ref{sec:2}. Next, in section \ref{sec4} we explore a few example scenarios whose signatures could be within reach of upcoming experiments. In this context, we establish a relationship between the tilt of the GW spectrum and its expected chirality. Finally, we show that our results and approximations are valid already in the small back-reaction regime. We offer our conclusions in \ref{sec:con}. We present some technical details on the normalization of the modes in the improved WKB scheme in appendix \ref{AppA} and expand on the fully numerical computation in appendix \ref{AppB}.

\section{Spectator axion-gauge field dynamics during inflation}
\label{sec:2}

In this section, we will briefly review the dynamics of the axion and gauge field sectors focusing on the vector field production resulting from the  rolling of the axion. Our discussion will be based on and will build upon the models introduced in \cite{Namba:2015gja} and \cite{Ozsoy:2020ccy} which we will follow closely.      

We consider an inflationary Lagrangian with a hidden axion sector, ${\rm U}(1)$ gauge fields, along with a canonical inflaton sector, all minimally coupled to gravity \cite{Barnaby:2012xt}:
\beq\label{S}
S=\int \mathrm{d}^4 x \sqrt{-g}\bigg[\frac{M_{\mathrm{pl}}^2 R}{2}-\underbrace{\frac{1}{2}(\partial \phi)^2-V_\phi(\phi)}_{\equiv\, \mathcal{L}_{\textrm{inflaton}}}-\underbrace{\frac{1}{2}(\partial \chi)^2-V_\chi(\chi)-\frac{1}{4} F_{\mu \nu} F^{\mu \nu}+\mathcal{L}_{\text {int }}}_{\equiv\, \mathcal{L}_{\textrm{hidden}}}\bigg],
\eeq
where $\phi$ is the inflaton and the spectator sector contains the axion-like particle $\chi$. The dynamics of the Abelian vector field $A_\mu$ is described by the anti-symmetric gauge field strength tensor $F_{\mu\nu} \equiv \partial_\mu A_\nu - \partial_\nu A_\mu$. The axion
enjoys an approximate shift symmetry and  couples to the vectors through the leading-order dimension 5 operator which takes the  form  
\beq\label{Lint}
\mathcal{L}_{\rm int} = - \frac{\lambda\, \chi}{8 f \sqrt{-g}} \,\epsilon^{\mu \nu \rho \sigma} F_{\mu \nu} F_{\rho \sigma},
\eeq
where $f$ is the axion decay constant, $\lambda$ is a dimensionless coupling constant and $\epsilon^{\mu\nu\rho\sigma}$ is the totally anti-symmetric symbol normalized as $\epsilon^{0123} = 1$. 
\medskip

\subsection{Background dynamics}\label{sec2p1}
For the background evolution of the system we consider a regime where the hidden (spectator) sector fields provide only a sub-dominant contribution to the total energy density during inflation.  This implies that (i) scalar field energy densities obey the hierarchy $\rho_\phi \gg \rho_\chi$, where $\rho_X = \dot{X}^2/2 + V_X(X)$, $X = \{\phi, \chi\}$, and (ii) gauge field fluctuations have no influence of the time evolution of the background $\rho_A \ll \rho_\chi$. Under these assumptions, and adopting the standard slow-roll hierarchy in the inflaton sector $\dot{\phi}^2/2 \ll V_\phi(\phi)$, the accelerated expansion is completely driven by the inflaton potential:
\beq
3H^2\Mp^2 = \rho_\phi + \rho_\chi + \rho_A \quad \longrightarrow \quad 3H^2\Mp^2 \simeq V_\phi(\phi).
\eeq
A sufficiently flat $V_\phi$ can then support a long quasi-dS expansion. In what follows we will leave the potential $V_\phi$ unspecified as the details regarding the inflaton sector dynamics are rather secondary in determining  our results. Working at leading order in the slow-roll expansion, we will treat the Hubble rate $H$ as constant and denote the scale factor during inflation in conformal time as $a(\tau) = -1/(H\tau)$, with $-\infty < \tau \leq 0$. 

Crucially, if the axion $\chi$ is displaced from its global minimum it can be dynamically active during inflation. In particular, we expect the potential $U_\chi$ to be sufficiently flat to allow slow-roll dynamics. In this work we will focus on two different axion potentials  \cite{Namba:2015gja,Ozsoy:2020ccy}:
\beq\label{pots}
U_\chi(\chi)=
 \begin{dcases} 
       \Lambda^4 \left[1-\cos\left(\frac{\chi}{f}\right)\right],& \quad ({\rm M}1) \,,\\
        \mu^3\chi + \Lambda^4 \left[1-\cos\left(\frac{\chi}{f}\right)\right]\,\,\,\,{\rm and}\,\,\,\, \Lambda^4\lesssim \mu^3 f&  \quad({\rm M}2),
   \end{dcases}
\eeq
where $\mu$ and $\Lambda$ are parameters of mass-dimension one.
\begin{figure}
\begin{center}
\includegraphics[scale=0.54]{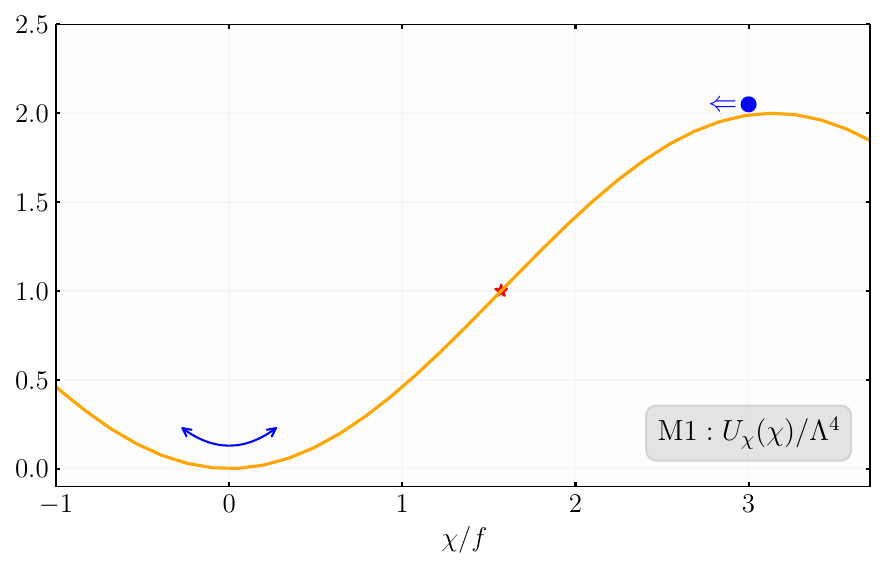}\includegraphics[scale=0.54]{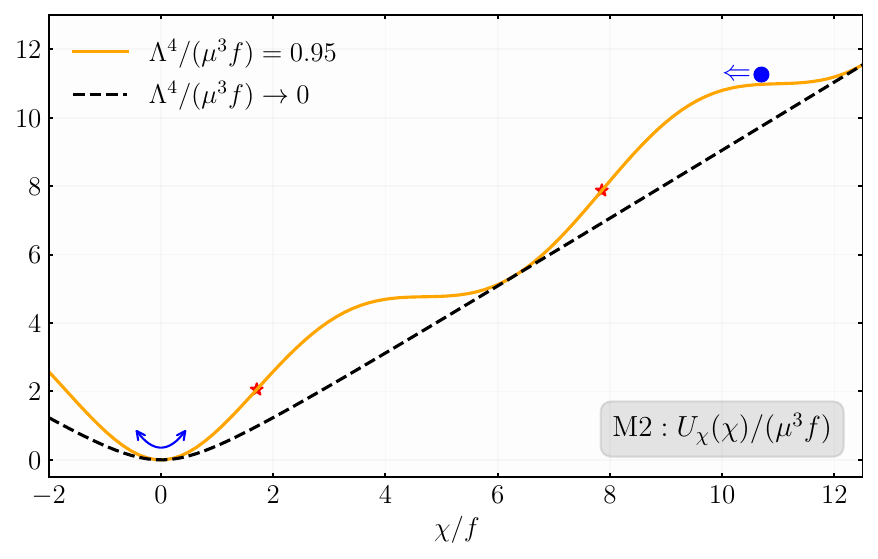}
\end{center}
\vspace*{-5mm}\caption{The shape of spectator axion potentials for M1 (left) and M2 (right). In both panels the red stars indicate the location of the inflections point(s) at which the slope of the potential $U'(\chi)$ and hence the background velocity $\dot{{\chi}}$ becomes maximal.\label{fig:POTS}}
\end{figure}
The first model, {\bf M1}, features a potential with a discrete shift symmetry just as in natural inflation \cite{Freese:1990rb}, where the size of the axion modulations is set by $\Lambda$. In this scenario, the motion of the axion takes place in between the  maximum ($\chi = \pi f$) and the minimum ($\chi = 0$) of the potential, so that for large  and small field values (early and late times) the axion rolls with small velocities. However, $\dot{\chi}$ is relatively large at intermediate times such as  when $\chi$ passes through an inflection point $\chi_* = \chi(t_*)$ with $U_\chi''(\chi_*) = 0$, where the slope of the potential $U_\chi'(\chi_*)$ becomes maximal. 

In the second model, {\bf M2}, the axion field range is extended via a linear term \cite{McAllister:2008hb,McAllister:2014mpa} proportional to a soft symmetry breaking mass parameter $\mu$. We assume that  $\chi$ can scan the corresponding bumpy potential\footnote{In this work, by an appropriate choice of initial conditions and model parameters, we assume that $\chi$ traverses two step-like features on its potential before it settles to its global minimum.} in the $\Lambda^{4}\lesssim \mu^3 f$ regime. In the plateau-like region(s), and towards the global minimum ($\chi = 0$) \footnote{The roll of $\chi$ towards the global minimum can be captured by modifying the monomial term as $\mu^3 \chi \to \mu^3 f [\sqrt{1 + (\chi/f)^2} -1 ]$, so that the axion potential \eqref{pots} interpolates between $\mu^3 \chi$ and $(\mu^3/f) \chi^2$ from large to small field ($\chi/f \to 0$) values respectively.}, the axion has very small velocity whilst it exhibits larger values when the slope of the potential $U'(\chi)$ becomes maximal at the cliff-like region(s), in particular at the inflection point(s) denoted by $^*$ in Fig.~\ref{fig:POTS}.

In the presence of the interaction in Eq.~\eqref{Lint}, the rolling axion  can have a substantial influence on the behaviour of gauge field fluctuations during inflation. We will review these effects below and point here to the extensive literature on the subject \cite{Namba:2015gja,Peloso:2016gqs,Garcia-Bellido:2016dkw,Garcia-Bellido:2017aan,Ozsoy:2017blg,Ozsoy:2020ccy,Ozsoy:2021onx,Fujita:2023inz,Unal:2023srk,Dimastrogiovanni:2023juq}. We will now expand on the dynamics of the vector fluctuations in the presence of the Chern-Simons coupling in Eq.~\eqref{Lint} and discuss the details of the particle production induced by the axion rolling down the potentials in Eq.~\eqref{pots}.

\subsection{ Revisiting vector field production by rolling spectator axions}\label{sec2p2}

{A dynamical axion $\chi(t)$ makes the CS term in Eq.~\eqref{Lint} important in that in this case the interaction can no longer be treated as a total derivative contribution, but rather as a consequential term in the Lagrangian  $\mathcal{L}_{\rm int}\propto \epsilon^{\mu\nu\rho\sigma} \chi(t) \partial_\mu (A_\nu (\partial_\rho A_\sigma))$. In particular, provided that the axion obtains a non-trivial velocity profile $\dot{\chi}\neq 0$, the dispersion relation of the gauge fields $A_\mu$ is modified leading to a copious production of gauge quanta. To see this, we decompose the gauge field into its Fourier modes \cite{Anber:2009ua}
\beq\label{Adc}
\hat{A}_i(\tau, {\bf x})= \int \frac{\mathrm{d}^3 {\bf k}}{(2 \pi)^{3 / 2}}\,\mathrm{e}^{i {\bf k} \cdot {\bf x}} \sum_{\lambda= \pm} \epsilon_i^{(\lambda)}({\bf k})\, \hat{A}_\lambda(\tau, \vbf{k}),
\eeq
where 
\beq
\hat{A}_{\lambda}(\tau, \vbf{k}) = A_\lambda(\tau, k)\,\hat{a}_\lambda(\vbf{k})+A_\lambda^*(\tau, k)\, \hat{a}_\lambda(-\vbf{k})^{\dagger}
\eeq
and the polarization vectors ($\lambda = \pm$) obey
\begin{align}\label{pve}
\nn & {k}_i\, \epsilon_i^{(\pm)}(\vbf{k}) =0, \quad \epsilon_{i j k}\, k_j \,\epsilon_k^{(\pm)}(\vbf{k})=\mp i|\vbf{k}|  \, \epsilon_i^{(\pm)}(\vbf{k}), \\ &\epsilon_i^{(\lambda)}(\vbf{k})\, \epsilon_i^{(\lambda^{\prime})}(\vbf{k})^* =\delta^{\lambda \lambda^{\prime}},\quad \epsilon_i^{(\pm)}(\vbf{k})^*=\epsilon_i^{(\pm)}(-\vbf{k})=\epsilon_i^{(\mp)}(\vbf{k}),
\end{align}
together with the commutation relations $[\hat{a}_\lambda(\vbf{k}),\hat{a}_{\lambda'}^{\dagger}(\vbf{k}')] = \delta_{\lambda \lambda'}\,\delta(\vbf{k}-\vbf{k}')$.
Using the decomposition \eqref{Adc} in the action \eqref{S}, the mode functions of the vector field $A_{\pm}$ satisfy
\beq\label{meqgf}
\partial_x^2 A_{\pm}+\left(1 \pm \frac{2 \xi}{x}\right) A_{\pm}=0, \quad\quad\quad\quad \xi \equiv-\frac{\lambda\, \dot{{\chi}}}{2 H f}\; ,
\eeq
where we defined a dimensionless time variable $x \equiv-k \tau = k/(aH)$, as well as  the effective dimensionless coupling $\xi$ between the spectator axion and the gauge field. Without  loss of generality, we work within the conditions $\xi\,>\,0$ and $\dot{\chi}<0$.

Notice from \eqref{meqgf} that the dimension five operator in \eqref{Lint} introduces a time dependent mass in the dispersion relation of gauge modes, which changes sign depending on the sign of the polarization $\lambda = \pm$. Deep inside the horizon, the modes satisfying the $x = k/(aH) \gg 1$  condition remain in their vacuum configuration as the time dependent mass in negligible. However, as the modes stretch outside the horizon, the $\lambda = -$ modes acquire a tachyonic mass for $ x \lesssim 2\xi$ and become amplified. It is clear from \eqref{meqgf} that the production of gauge quanta is also sensitive to the velocity profile of the axion-like field  which in general makes $\xi$ time dependent.  Thanks to the relatively simple form of the potentials in \eqref{pots}, analytic expressions for the velocity profile $\dot{\chi}$ and hence for the effective coupling $\xi$ can be derived in the slow-roll regime, $\ddot{\chi} \ll 3H\dot{\chi}$. For typical field ranges\footnote{This implies a single inflection point for the $\textrm{M}1$ model and one  cliff-like region for $\textrm{M}2$.} dictated by the potentials in \eqref{pots} $\xi$ displays the following peaked profiles \cite{Namba:2015gja, Ozsoy:2020ccy}:
\beq\label{xip}
\xi(x)= \begin{dcases}\frac{2 \xi_*}{\left(x_* / x\right)^\delta+\left(x / x_*\right)^\delta}, & \delta \equiv \frac{\Lambda^4}{3 H^2 f^2}\,\,\, {\rm and} \,\,\,\xi_*\equiv \frac{\lambda\, \delta}{2},\,\, ({\rm M}1), \\ \frac{\xi_*}{1+\delta^2\ln\left(x / x_*\right)^2}, & \delta \equiv \frac{\mu^3}{3 H^2 f} \,\,\, {\rm and} \,\,\,\xi_*\equiv {\lambda\, \delta},\,\, ({\rm M}2),\end{dcases}
\eeq
where $x_* \equiv -k \tau_*$ is the time when the axion passes through the inflection point and $\xi_*$ is the maximum value of the effective coupling at that point. The width of the time dependent peak in $\xi$ is controlled by the dimensionless ratio $\delta$ which roughly characterises the mass of the axion at its global minimum, $\delta \approx m_\chi^2 / H^2$. It follows that for a heavier axion (corresponding to larger $\delta$) the restoring force towards the global minimum is larger and the axion traverses the inflection points faster, leading to sharper peak(s) in $\xi$. From this perspective, $\delta$ can be considered as a measure of the axion's acceleration, $\dot{\xi}/(\xi H) \simeq \ddot{\chi}/(\dot{\chi} H) \sim \delta$ as it rolls down on its potential. Note that consistency with the slow-roll approximation employed in deriving \eqref{xip}  requires that we limit our analysis to the $\delta \ll 3$ regime. We will adopt the value $\delta = 0.4$ when presenting our results. 

Using the analytic forms \eqref{xip} for the effective coupling in Eq.~\eqref{meqgf} does not allow one to obtain a closed form analytic solution for the vector modes for a general set of model parameter $\{\delta, \xi_*, x_*\}$ that parametrize the background evolution of the axion. This limitation notwithstanding, a solution for the enhanced vector mode $A_{-}$ can be found using semi-analytic methods including the WKB approximation. The approximate solution models the time dependence of the axion velocity  around the inflection points by means of a scale-dependent amplification of gauge modes in terms of a normalization factor which depends on the model parameters $N = N(\xi_*, x_*, \delta)$.
In terms of $N$ the growing solution describing the amplification in the vector mode functions can be written as 
\beq\label{Amsol}
A_{-}(\tau, k) \simeq \frac{N\left(\xi_*,-k \tau_*, \delta\right)}{\sqrt{2 k}}\left[\frac{-k \tau}{2 \xi(\tau)}\right]^{1 / 4} \exp \left[-2E(\tau) \sqrt{-2 \xi_* k \tau_*}\right], \quad\quad \tau / \tau_* < 1,
\eeq
where $E(\tau)$ is a model dependent function that asymptotes to zero at late times ($\tau \to 0$):
\beq\label{Eform}
E(\tau)= \begin{dcases}\frac{\sqrt{2} \left(\tau / \tau_*\right)^{(1+\delta) / 2}}{(1+\delta)}, & (\text{M1}) \\ -\frac{\rm{Ei}\left[\,{\ln(\tau/\tau_*)}/2\,\right]}{2\delta}, &  (\text{M2), }
\end{dcases}
\eeq
where $\rm{Ei}$ is the exponential integral\footnote{Note that for the M2 model, previous works (see \eg \cite{Ozsoy:2020ccy,Ozsoy:2021onx,Ozsoy:2023ryl}) use a different functional form for the argument $E(\tau)$  of the exponential in the gauge field mode function \eqref{Amsol}. In fact, the expression used for the argument in those works corresponds to the asymptotic  $\tau/\tau_* \to 0$ limit of the expression we use in the second line of \eqref{Eform}. In light of our findings for the fully numerical approach detailed in Appendix \ref{AppA}, we also conclude that our approximate solution leads to more accurate results for the observables under scrutiny (see also section \ref{sec4}.).}.
Note that the scale dependence of the solution \eqref{Amsol} is a direct consequence of the time dependent mass of the mode functions which becomes maximally tachyonic at $\tau = \tau_*$ (at the inflection points) for modes that have a size roughly equal to the horizon size at this time, \ie $x_* \sim \mathcal{O}(1)$. These modes exhibit the highest amplitude $N = N(\xi_*, x_*, \delta)$ as compared to the modes that satisfy $x_* \gg 1$ and $x_* \ll 1$. 

In order to obtain the normalization factor, we work with fixed values of $\xi_*$ and $\delta$ and numerically evaluate the mode equation of the vector fields in \eqref{meqgf} using \eqref{xip} for a grid of $x_*$ values and match it to the WKB solution \eqref{Amsol} (see Appendix \ref{AppA}). A common practice at this stage is to fit the resulting $N = N(\xi_*, x_*, \delta)$ to a nice analytical log-normal form that was first introduced in \cite{Namba:2015gja}: 
\beq\label{Nform}
N\left(\xi_{*}, x_*, \delta\right) \simeq N^{c}\left[\xi_{*}, \delta\right] \exp \left(-\frac{1}{2 \sigma^{2}\left[\xi_{*}, \delta\right]} \ln ^{2}\left(\frac{x_*}{q^{c}\left[\xi_{*}, \delta\right]}\right)\right),
\eeq
where in this approximation, $N^c$, $q^c$ and $\sigma$ describe the peak amplitude, peak location and width of the produced gauge quanta. Using \eqref{Nform}, one can obtain accurate fitting functions in terms of $\xi_*$ at fixed $\delta$. 
\begin{table}
\begin{center}
\begin{tabular}{|c |c |c |c |}
\hline
\hline
\cellcolor[gray]{0.9}&\cellcolor[gray]{0.9}$\ln(N^{c})\simeq$&\cellcolor[gray]{0.9}$q^{c}\simeq$&\cellcolor[gray]{0.9}$\sgm\simeq$ \\
\hline
\cellcolor[gray]{0.9}\scalebox{0.95}{${{\rm M}1}$}&\scalebox{0.9}{$0.183 + 2.69\,\xi_*+ 0.0007\,\xi_*^2$}&\scalebox{0.9}{$-0.062 +0.660\, \xi_* - 0.0007\, \xi_* ^2$}&\scalebox{0.9}{$1.67 -0.234\, \xi_* + 0.0144\, \xi_* ^2$}\\\hline
\cellcolor[gray]{0.9}\scalebox{0.95}{${{\rm M}2}$}&\scalebox{0.9}{$ 0.234+2.55 \,\xi_{*}-0.0010 \,\xi_{*}^{2}$} & \scalebox{0.9}{$0.030 + 0.743 \,\xi_{*}-0.0007\, \xi_{*}^{2}$}&\scalebox{0.9}{$1.38-0.193\, \xi_{*} + 0.0118\, \xi_{*}^{2}$} \\\hline
\hline
\end{tabular}
\caption{\label{tab:Nfitapp} $\xi_*$ dependence of the peak amplitude $N^{c}$, peak location $q^c$ and width $\sgm$ that parametrize the approximate log-normal form (eq. \eqref{Nform}) of the gauge field normalization factor. We pick $\delta = 0.4$ and $3\leq \xi_*\leq 6.5$.}
\end{center}			
\end{table}
\begin{figure}[t!]
\begin{center}
\includegraphics[scale=0.56]{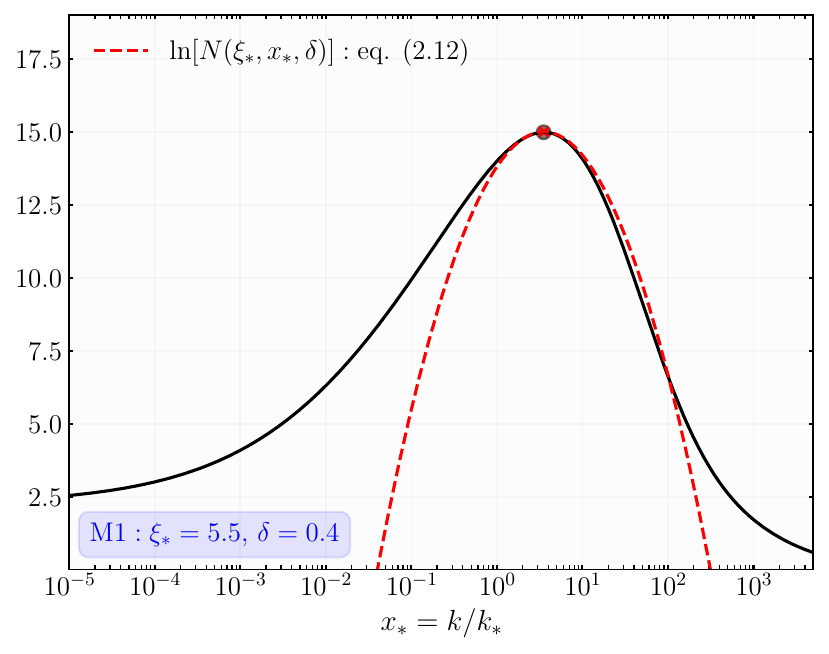}\includegraphics[scale=0.56]{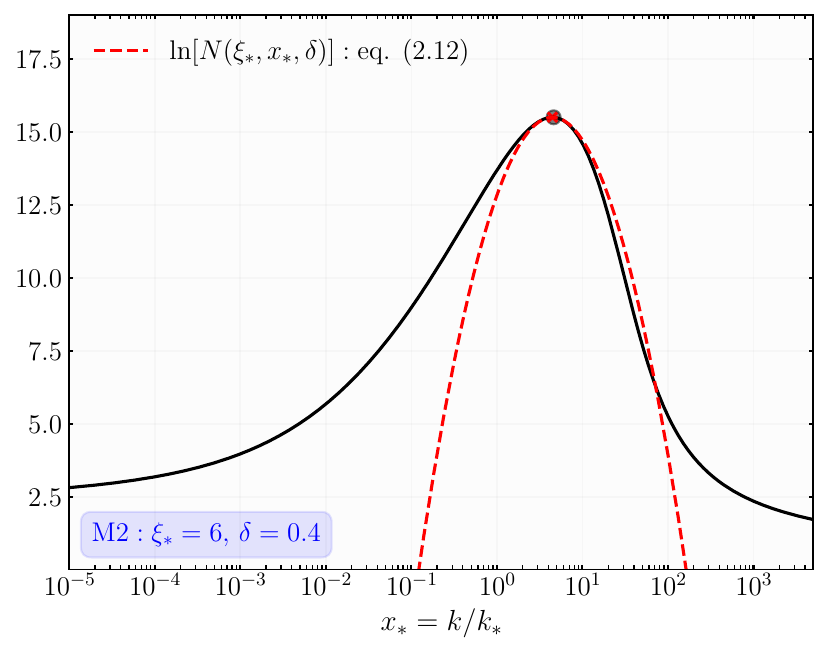}
\end{center}
\vspace*{-5mm}\caption{A comparison of the normalization factors, $\ln[N(\xi_*, x_*, \delta)]$, obtained with the WKB approximation of Eq.~\eqref{Amsol}.  The continuous black curve corresponds to the numerically obtained $N$. Red dashed lines account for another layer of approximation, namely that of Eq.~\eqref{Nform}.
\label{fig:norms}}
\end{figure}

Although this approach provides a convenient analytic form to parametrize the scale dependence of the enhanced mode functions, it does not fully capture the IR and UV behavior of the normalization function. To illustrate this, we plot in Fig.~\ref{fig:norms} a numerical fit for the normalization factor $N$ \textit{vs} the approximate log-normal form in Eq.~\eqref{Nform} (see Table \ref{tab:Nfitapp}). Both the numerical and analytical fit are used within the WKB approximation scheme. It is clearly visible from the  plots in Fig.~\ref{fig:norms}  that the scale dependence obtained by means of the two different approaches is very similar in the vicinity of the maximum (marked with a black dot). However, away from the peak  the two methods lead to considerably different profile for the normalization factor. In particular, $N$ and, in turn, the amplitude of the gauge fields carry much more power in the IR ($x_* \ll q^c$) when the numerical route to calculate $N$ is used as opposed to the standard analytic approach of Eq.~\eqref{Nform}.

The ``IR/UV'' asymmetry visible in Fig.~\ref{fig:norms} for the vector field amplitude stems from the fact that soft modes spend more time in the tachyonic regime compared to modes that satisfy $x_* = k/k_* \gg 1$ as they exit the horizon. This is in complete analogy with the enhancement of scalar entropy modes in scenarios comprising turns in field space during inflation \cite{Fumagalli:2020adf}. The consequences on cosmological correlators stemming from the above findings are at the heart of the present work. Vector fields, amplified by the rolling axion, act as sources for the 1-loop scalar and tensor power spectra. It follows that the precise momentum dependence of the gauge quanta wave-function  is crucial in accurately deriving the cosmological signal. Motivated by such considerations, we will  go beyond the WKB approximation undertaken in this section and provide a fully numeric approach, using the techniques of \cite{Garcia-Bellido:2023ser}, to compute the mode evolution of the vector fields (see Appendix \ref{AppA}) and the resulting cosmological signals (see Appendix \ref{AppB}).  
\begin{figure}[t!]
\begin{center}
\includegraphics[scale=0.6]{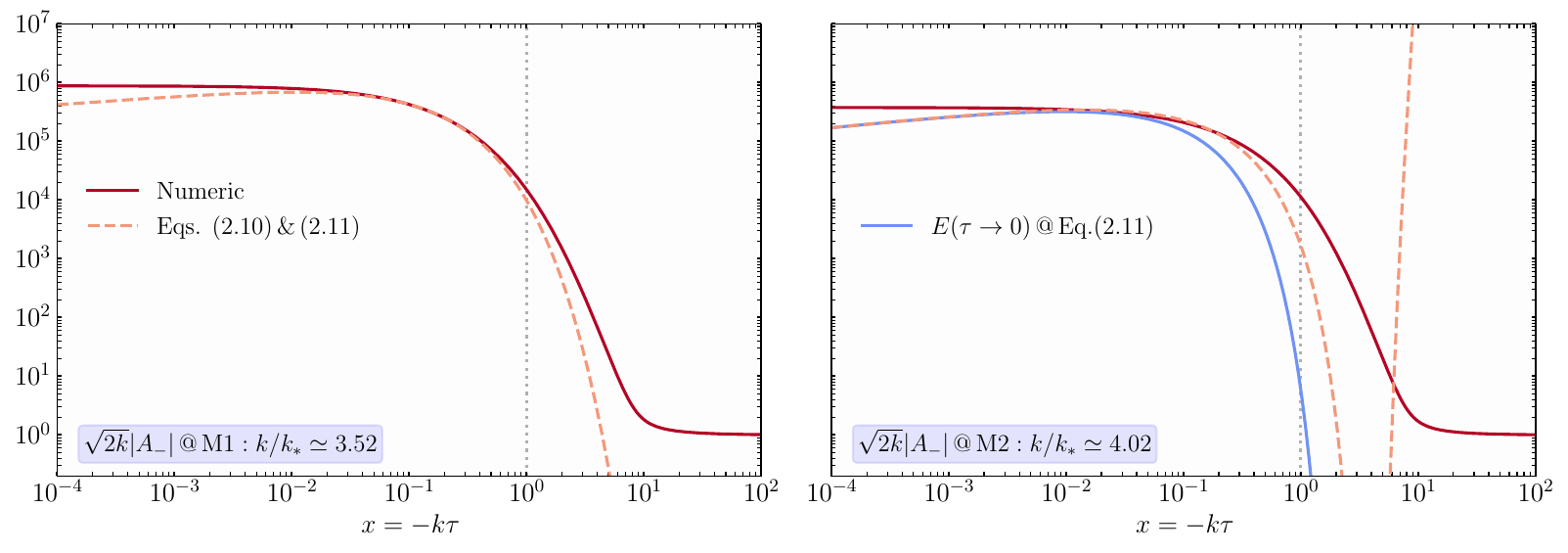}
\end{center}
\vspace*{-5mm}\caption{Time evolution (right to left) of the vector field mode functions obtained by numerical evaluation (solid) and WKB solution (dashed) using numerically calculated normalization function $N(\xi_*,k/k_*,\delta)$ in \eqref{Amsol}. The light blue curve for the M2 model is obtained via the WKB solution using the asymptotic limit $\tau \to 0$ of \eqref{Eform} corresponding to the solution used in \cite{Ozsoy:2020ccy,Ozsoy:2021onx,Campeti:2022acx}. In both plots we take $\xi_* = 5.5$ and $\delta = 0.4$.\label{fig:vmf}}
\end{figure}

Before moving on to the observables, we find it useful to compare the accuracy of the WKB solution for the gauge wave-function in Eq.~\eqref{Amsol}  to the fully numerical solution obtained by solving Eqs. \eqref{meqgf} and \eqref{xip} (see also Appendix \ref{AppA}). In Fig.~\ref{fig:vmf} we plot the evolution of the maximally enhanced vector modes. The  amplification of the ${\rm U}(1)$ field -- manifest already inside the horizon for $x \lesssim \mathcal{O}(10)$ -- is shown by the solid lines. The solution obtained via the WKB scheme (see Eqs. \eqref{Amsol} and \eqref{Eform}) with the numerically obtained normalization $N$ is represented by the dashed lines. We observe that for the M1 model, the WKB solution captures the particle production effects in the gauge sector fairly well for the domain where most of the amplifications takes place $10^{-2} \lesssim x \lesssim \mathcal{O}(1)$. On the other hand, the WKB solution in the M2 model performs worse within the same time domain\footnote{The sudden growth for the WKB solution in M2 is unphysical. This is remedied  by introducing a cut-off procedure when computing cosmological correlators sourced by  vector fields (see Appendix \ref{AppB}).}.

Both tensor and scalar fluctuations are sourced non-linearly by gauge fields so that one can expect the ability of the WKB approximation to approach the fully numerical result in the M1 \textit{vs} M2 model to be reflected also at the level of the power spectra (see section \ref{sec3}). As we will show (see section \ref{sec4}), the accuracy gap between the WKB and the fully numerical solutions will be more prominent in the case of the scalar spectrum. Scalar fluctuations are indeed more sensitive than tensors to the time evolution of the vector field in the key time domain where the bulk of the gauge modes production takes place (see Appendix \ref{AppB}).

\section{Revisiting scalar and tensor perturbations sourced by the gauge fields}\label{sec3}

In the spectator axion models we are considering, the impact of the vector field production on the visible scalar and tensor fluctuations arise through gravitational interactions \cite{Barnaby:2012xt, Ferreira:2014zia, Mukohyama:2014gba}. Below, we will briefly review below the impact of gauge fields on these fluctuations. 

\noindent{\bf Tensor perturbations.} To study the sourcing of tensor perturbations by amplified gauge fields, we focus on the transverse traceless metric perturbation  $g_{ij} = a^2(\tau) [\delta_{ij} + \hat{h}_{ij}(\tau,\vec{x})]$ and decompose it into circularly polarized states $\lambda = \pm$ in Fourier space as $\hat{h}_\lambda (\tau, \vec{k}) = \Pi_{ij,\lambda} (\vec{k})\,\hat{h}_{ij}(\tau,\vec{k})$ where $\Pi_{ij,\lambda}$ is the polarization tensor obeying $\hat{k}_i\, \Pi_{ij,\lambda}(\vec{k})=0 $,  $\Pi^{*}_{ij,\lambda}\Pi_{ij,\lambda'} = \delta_{\lambda\lambda'}$ and $\Pi^{*}_{ij,\lambda}(\vec{k}) = \Pi_{ij,-\lambda}(\vec{k}) = \Pi_{ij,\lambda}(-\vec{k})$. Expanding the action \eqref{S} up to third order in gauge field and metric fluctuations, one obtains for the graviton polarization states $h_{\lambda}$ \cite{Barnaby:2012xt}:
\begin{equation}\label{te}
\left(\partial^2_\tau + k^2 -\frac{2}{\tau^2}\right)(a \hat{h}_\lambda) =-\frac{2a^{3}}{\Mp^2} \Pi_{i j, \lambda}(\vec{k}) \int \frac{\d^{3} x}{(2 \pi)^{3 / 2}} \mathrm{e}^{-i \vec{k} \cdot \vec{x}}\left[\hat{E}_{i} \hat{E}_{j}+\hat{B}_{i} \hat{B}_{j}\right]
\end{equation}
where we defined dark ``electric" and ``magnetic" fields $\hat{E}_i (\tau, \vec{x}) = -{a^{-2}} \hat{A}_{i}^{\prime}(\tau, \vec{x}), \,\, \hat{B}_{i}={a^{-2}} \epsilon_{i j k} \partial_{j} \hat{A}_{k}$ in a flat FLRW universe, in analogy with standard electromagnetism.  
\smallskip

\noindent{\bf Scalar perturbations.} The role of particle production in sourcing  the (scalar) curvature perturbation $\zeta$ is indirect and mediated by the spectator axion.

The action \eqref{S} does not include direct couplings between the axion and the inflaton $\phi$. However, integrating out the non-dynamical scalar metric fluctuations, namely the lapse $\delta N$ and the shift $N^{i}$ (i.e. solving the constraint equations), results in a mass mixing between the inflaton $\delta \phi$ and spectator axion $\delta \chi$ fluctuations. It follows then that gauge fields affect curvature fluctuations\footnote{Note that in principle  $\zeta$ could also obtain a direct contributions from  $\delta \chi$ and gauge quanta. In this work we will consider a spectator axion that rolls down to its minimum long before the end of inflation (see section \ref{sec2p1}) so that the contribution of $\delta \chi$ to $\zeta$ is negligible at late times  \cite{Mukohyama:2014gba,Namba:2015gja}. The contribution from gauge fields is roughly proportional to the absolute value of Poynting vector, $a|\vec{S}| = a|\vec{E}\times \vec{B}|$ which is also negligible at late times when the particle production saturates at super-horizon scales and the resulting electromagnetic fields decay as $\vec{E},\vec{B} \sim a^{-2}$ \cite{Ozsoy:2020ccy}. One can therefore safely employ 
the standard relation $\zeta = - H \delta \phi/\dot{\phi}$ in what follows.}  via $A^2 \to \delta \chi \to \delta \phi \to \zeta$. Let us now  focus on how gauge fields source the axion scalar fluctuation $\delta \chi$,
\begin{equation}
\label{usgm}\left(\frac{\partial^{2}}{\partial \tau^{2}}+k^{2}-\frac{2}{\tau^{2}}\right) (a \delta \hat{\chi}) \simeq \frac{a^{3}\lambda}{f} \int \frac{\d^{3} x}{(2 \pi)^{3 / 2}} \mathrm{e}^{-i \vec{k} \cdot \vec{x}} \, \hat{E}_i(\tau, \vec{x}) \hat{B}_i(\tau, \vec{x}).
\end{equation}
Equipped with the inhomogeneous solution for $\delta \chi$  in \eqref{usgm}, one then arrives at  $\delta \phi$ via the relation
\beq
\label{uphi} \left(\frac{\partial^{2}}{\partial \tau^{2}}+k^{2}-\frac{2}{\tau^{2}}\right) (a \delta \hat{\phi}) \simeq 3a^2 \frac{\dot{\phi}\dot{\chi}}{\Mp^2} ~(a \delta \hat{\chi}),
\eeq
which has also been obtained after solving the constraint equations. The particular solutions to Eq.~\eqref{te} and Eqs.~ \eqref{usgm}-\eqref{uphi} account for an additional source of tensor and scalar fluctuations besides the vacuum fluctuations  \ie those corresponding to the homogeneous solution to Eqs.~\eqref{te} and \eqref{uphi}. Sourced and vacuum contributions are statistically uncorrelated and hence can be simply added when calculating key observables such as the total scalar and tensor power spectra.

\subsection{The power spectra of cosmological perturbations}

We use $\hat{\mathcal{X}}$ as a placeholder for curvature and GW fluctuations, $\hat{\mathcal{X}} = \{\hat{\zeta},\hat{h}_\pm\}$. Each will receive a vacuum and a sourced contribution so that $\hat{\mathcal{X}} = \hat{\mathcal{X}}^{(\rm v)} + \hat{\mathcal{X}}^{(\rm s)}$. We define the corresponding power spectra as 
\begin{align}\label{psdef}
\nn \langle\hat{\zeta}(\vbf{k}) \hat{\zeta}(\vbf{k'})\rangle &\equiv  \,\delta(\vbf{k}+\vbf{k}') \,
\frac{2\pi^{2}}{k^{3}}\mathcal{P}_{\zeta}\left(k\right) \;,\\
\langle \hat{h}_{\lambda}(\vbf{k}) \hat{h}_{\lambda'} (\vbf{k'})\rangle & \equiv \, \delta_{\lambda \lambda'}\, \delta(\vbf{k}+\vbf{k}')\, \frac{2 \pi^{2}}{k^{3}} \mathcal{P}_{\lambda}({k}). 
\end{align}
The origin of the vacuum and sourced part of perturbations are different and so they are statistically uncorrelated. The total power spectra can then be simply obtained by the sum of vacuum and sourced auto-correlators as 
\begin{equation}\label{tps}
\mathcal{P}_{\zeta}(k) = \mathcal{P}^{(\rm v)}_{\zeta}(k) + \mathcal{P}^{(\rm s)}_{\zeta}(k),\quad\quad \mathcal{P}_{h}(k) = \sum_{\lambda = \pm} \mathcal{P}_{\lambda} (k) = \sum_{\lambda = \pm}  \mathcal{P}^{(\rm v)}_{\lambda}(k) + \mathcal{P}^{(\rm s)}_{\lambda}(k)
\end{equation}
where the vacuum contributions are given by the standard expressions:
\begin{equation}\label{psv}
\mathcal{P}_{\zeta}^{(\rm v)} = \fr{H^2}{8\pi^2\epsilon_\phi \Mp^2},\quad\quad \mathcal{P}^{(\rm v)}_{h} \equiv \sum_{\lambda = \pm}  \mathcal{P}^{(\rm v)}_{\lambda}(k) = \fr{2 H^2}{\pi^2\Mp^2} ,
\end{equation}
and we defined $\epsilon_\phi \equiv \dot{\phi}^2/ (2 H^2 \Mp^2)$ as the slow-roll parameter controlled by the inflaton sector. 
\begin{table}[t!]
\begin{center}
\begin{tabular}{| c | c | c | c |}
\hline
\cellcolor[gray]{0.9}$\{i,j\}$&\cellcolor[gray]{0.9}$\ln(f^c_{i,j}) \simeq $&\cellcolor[gray]{0.9}$x^c_{i,j} \simeq$&\cellcolor[gray]{0.9}$\sgm_{i,j} \simeq$\\
\hline
$\cellcolor[gray]{0.9}\{2,-\}$ & \scalebox{0.95}{$-7.56 +9.50\, \xi_* + 0.0652\, \xi_* ^2$}&\scalebox{0.95}{$2.37 + 0.949\,\xi_* + 0.0106\, \xi_*^2$}&\scalebox{0.96}{$0.78 -0.091\,\xi_* + 0.0054\,\xi_* ^2 $}\\\hline
$\cellcolor[gray]{0.9}\{2,+\}$ & \scalebox{0.95}{$-13.33 +9.43\, \xi_* + 0.0705\, \xi_* ^2$ }&\scalebox{0.95}{$1.03 + 0.425\,\xi_* + 0.0058\, \xi_*^2$}&\scalebox{0.95}{$0.87 - 0.074\,\xi_* + 0.0045\,\xi_* ^2 $}\\\hline
$\cellcolor[gray]{0.9}\{2,\zeta\}$&\scalebox{0.95}{$ -6.67 + 9.50\,\xi_*+ 0.0639\,\xi_*^2$}&\scalebox{0.95}{$1.44 + 0.662\, \xi_* + 0.0024\, \xi_* ^2$}&\scalebox{0.95}{$0.85 -0.087\, \xi_* + 0.0055\, \xi_* ^2$}\\\hline
\hline
\hline
$\cellcolor[gray]{0.9}\{2,-\}$ &\scalebox{0.95}{$ -9.89 + 9.47\,\xi_*+ 0.0254\,\xi_*^2$} & \scalebox{0.95}{$3.47 +0.805\, \xi_* + 0.0293\, \xi_* ^2$}&\scalebox{0.95}{$0.58 -0.058\, \xi_* + 0.0032\, \xi_* ^2$}\\\hline
$\cellcolor[gray]{0.9}\{2,+\}$ & \scalebox{0.95}{$-15.91 +9.46\, \xi_* +0.0267\, \xi_* ^2$} & \scalebox{0.95}{$1.48 + 0.396\,\xi_* + 0.0105\,\xi_* ^2$}& \scalebox{0.95}{$ 0.71 -0.046\,\xi_* + 0.0028\,\xi_* ^2$} \\\hline
$\cellcolor[gray]{0.9}\{2,\zeta\}$& \scalebox{0.95}{$ -10.16 + 9.70\,\xi_*+ 0.0120\,\xi_*^2$} & \scalebox{0.95}{$2.71 + 0.466\, \xi_* + 0.0210\, \xi_* ^2$} & \scalebox{0.95}{$0.65 - 0.050\, \xi_* + 0.0031\, \xi_* ^2$} \\\hline
\end{tabular}
\caption{\label{tab:fit1} $\xi_*$ dependence of peak height, width and the location of the Gaussian template \eqref{f2form} in the Models M1 (top three rows) and M2 (bottom three row) for $\delta = 0.4$ and ${3\leq \xi_*\leq 6.5}$.}
\end{center}			
\end{table}
The sourced power spectra in \eqref{tps} inherit the scale dependence of  gauge fields which in this context can be parametrized as (see Appendix \ref{AppB}) \cite{Namba:2015gja},
\begin{align}\label{SC}
\nn
\mathcal{P}^{(\rm s)}_{\zeta}(k) &=\left[\epsilon_{\phi} \mathcal{P}_{\zeta}^{(\rm v)}(k)\right]^{2} \sum_i f^{(i)}_{2, \zeta}\left(\xi_{*},\frac{k}{k_{*}}, \delta\right),\\
\mathcal{P}^{(\rm s)}_{\pm}(k) &=\frac{\left[\mathcal{P}_{h}^{(\rm v)}(k)\right]^{2}}{256} \sum_i f_{2, \pm}^{(i)}\left(\xi_{*},\frac{k}{k_{*}}, \delta\right),
\end{align}
where the summation over $i$ denotes contributions from each possible particle production site as the axion rolls down  its potential \eqref{pots} towards the global minimum  (see Fig. \ref{fig:POTS}). For the {\bf M1} model  the sum is over only  one such region, whilst  there are two of them in the {\bf M2} case (see section \ref{sec2p1}). The function $ f_{2}$ fully characterizes the effect of particle production and its underlying dependence on the dynamics of the spectator axion. In particular, for fixed axion acceleration $\delta$, the amplitude of the power spectra in \eqref{SC} is controlled by the maximal value of the effective coupling $\xi_*$ while the scale dependence $x_* = k/k_*$ depends on the underlying vector field fluctuations. 

We would now like to expand  on the two different approaches  discussed in section \eqref{sec2p2} for the computation of $f_{2,j}$. The standard practice for the computation  is to employ the approximate log-normal form of the vector field normalization within the WKB solution \eqref{Amsol}. This is done in order to fit the scale dependence of the resulting correlator for different effective coupling values $\xi_*$ (see \eg \cite{Namba:2015gja} and \cite{Ozsoy:2020ccy}). Such procedure results in a log-normal distribution for $f_{2,j}$ 
\beq\label{f2form}
f_{2, j}\left( \xi_{*},\frac{k}{k_{*}}, \delta\right) \simeq f_{2, j}^{c}\left[\xi_{*}, \delta\right] \exp \left[-\frac{1}{2 \sigma_{2, j}^{2}\left[\xi_{*}, \delta\right]} \ln ^{2}\left(\frac{k}{k_{*} x_{2, j}^{c}\left[\xi_{*}, \delta\right]}\right)\right].
\eeq
\begin{figure}[t!]
\begin{center}
\includegraphics[scale=0.57]{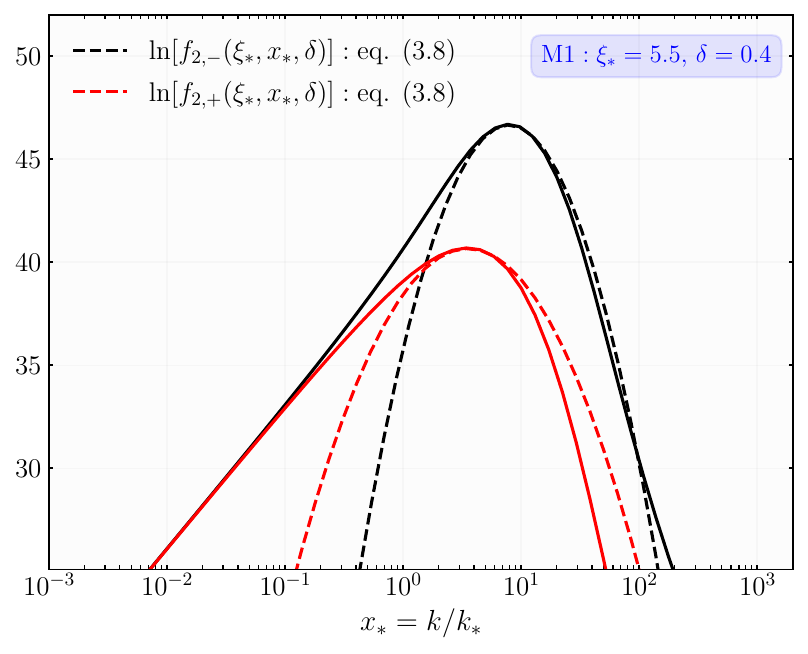}\includegraphics[scale=0.57]{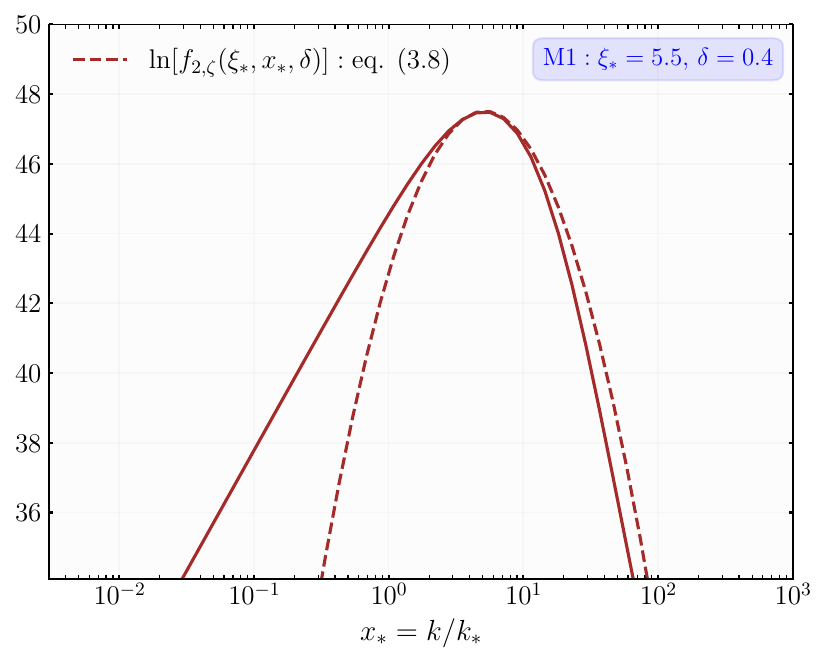}
\includegraphics[scale=0.58]{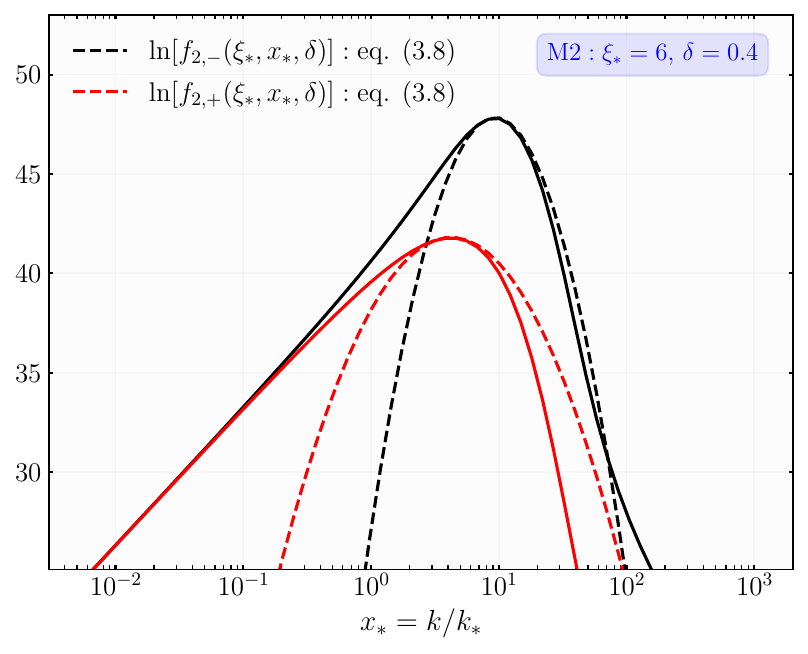}\includegraphics[scale=0.58]{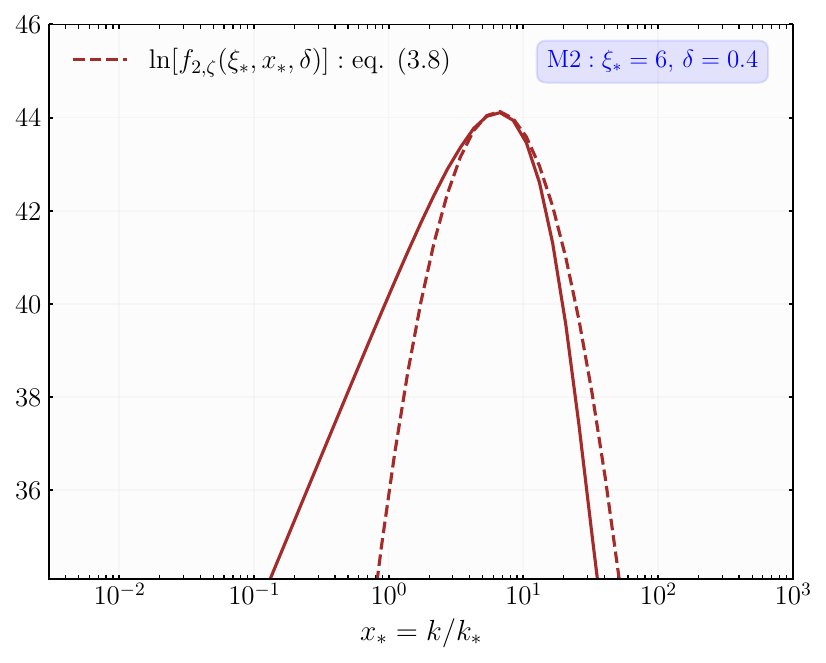}
\end{center}
\vspace*{-5mm}\caption{Sourced contributions $f_{2,j}$ (solid curves) for the power spectra of tensor $j = \pm$ (left) and scalar $j = \zeta$ (right) perturbations for the Model ${\rm M}1$ (top) and ${\rm M}2$ (bottom). Dashed curves represent the Gaussian fitting form in \eqref{f2form} where Table \ref{tab:fit1} is utilized.\label{fig:f2s}}
\end{figure}

\noindent However, as anticipated in section \ref{sec2p2}, this approach leads to inaccurate results since the true profile (see the black curve in Fig.~\ref{fig:norms}) of gauge field fluctuations is considerably different from the Gaussian approximation presented in \eqref{Nform}. In Fig.~\ref{fig:f2s} we underscore this fact by comparing the scale dependence of $f_{2,j}$ in both approaches using a representative choice of the model parameters. It is clear that, although the Gaussian form \eqref{f2form} fits the numerical result in the WKB approach very well in the vicinity of the peaks, it fails to accurately capture especially the IR behavior. Another important lesson we learn from Fig.~\ref{fig:f2s} is the suppression of the net chirality of tensor modes in the IR, where the solid curves in the left panels overlap. {This is a consequence of the fact that the phase space for the production of soft tensor modes is dominated by the enhanced gauge fields that exhibit relatively large loop momenta. As a result, large wavelength ($x_* = k/k_* \ll 1$) sourced tensor perturbations are comparable for the two GW polarization (left/right) states. This leads to the suppression of net chirality of the produced GWs in the IR, as first emphasized in \cite{Garcia-Bellido:2023ser}. For the sake of completeness and clarity, we have now added a brief technical discussion on these important points in Appendix B.}
The results derived in this section have potentially important phenomenological implications. First of all, if the rolling of the axion occurs sufficiently early  during inflation, it can generate an observable chiral GW background that dominates over the vacuum at CMB scales. This opens up the possibility of testing even a relatively low scale realization of inflation \cite{Namba:2015gja, Ozsoy:2020ccy}. However, the amplitude of the sourced GW signal is constrained by the well-established bounds on the scalar perturbations: these can be quite large given that they are sourced by the very same mechanism that generates tensor perturbations (see \eg \cite{Namba:2015gja,Campeti:2022acx}). It becomes then crucial to fully characterize sourced tensor perturbations at large scales focusing on observables sensitive to the chirality (\eg ${\rm TB, EB, BBB}$) of the signal \cite{Shiraishi:2016yun,LiteBIRD:2023zmo}, as well as the scale dependence and amplitude of scalar perturbations (\eg ${\rm TT, TTT, TBB, EBB, BTT, BEE}$)  \cite{Namba:2015gja, Ozsoy:2021onx}.

Interestingly, various axion-gauge field models have sufficient freedom in their field and parameter space to support striking features in the power spectra at intermediate and/or small scales. These corresponds to the bulk of the axion's displacement taking place during the late stages of inflation. The ensuing possibilities include the production of primordial black holes \cite{Ozsoy:2023ryl} along with the direct and indirect (\eg scalar-induced) production of a stochastic GW background at interferometer scales \cite{Garcia-Bellido:2016dkw, Garcia-Bellido:2017aan, LISACosmologyWorkingGroup:2023njw}. The amplitude and shape of these signals is naturally very sensitive to the underlying vector sources which, in turn, are regulated by the rolling axion dynamics. In the next section, we will focus on scalar and tensor perturbations relevant for PBH physics and on the generation of stochastic GW backgrounds at interferometer scales.

\section{Phenomenology at intermediate and small  scales}\label{sec4}
Armed with the results obtained for the 2-point cosmological correlators in the previous section, we are now in a position to dive into the phenomenological implications of the axion-gauge field dynamics during inflation.

In what follows we shall assume that the spectator axion reaches its maximal velocity (and, correspondingly, its leading effect on the observables) far after the time when CMB scales exit the horizon. It follows that  the dominant contribution to the scalar and tensor spectra at large scales will be due to vacuum fluctuations. Compliance with CMB observations \cite{Akrami:2018vks} then places constraints on the inflaton potential $V_\phi(\phi)$\footnote{See \eg \cite{Alam:2024krt} for a recent work in the context of natural inflation that exhibit direct CS type coupling with gauge fields.}. In this work we shall be agnostic about the explicit form of the potential whilst ensuring that the slow-roll parameters of the model lead to a scalar power spectrum and spectral tilt that are fully compatible with CMB measurements.

We take $r(k_{\rm cmb}) = 0.03$, close to the current limits by Planck and BICEP/Keck Array data \cite{Tristram:2021tvh,Campeti:2022vom}. Enforcing the consistency relation $r = 16 \epsilon_\phi = -8 n_t$ at CMB scales then allows us to fix the inflaton slow-roll parameter\footnote{Neglecting higher orders in slow-roll, we assume that $\epsilon_\phi$ remains fixed throughout inflation.} and the tensor tilt as $\epsilon_\phi = 1.875 \times 10^{-3}$ and $n_t = - 3.75 \times 10^{-3}$. The Planck results provide the scalar tensor tilt and the amplitude as $n_s - 1 \simeq 0.035$ and $\mathcal{P}_{\zeta}(k_{\rm cmb}) \equiv \mathcal{A}_s \simeq  2.1 \times 10^{-9}$ \cite{Akrami:2018vks}. Combining these bits of information provides the overall amplitude and scale dependence of the vacuum power spectra:\begin{align}\label{psvp}
\mathcal{P}^{(\rm v)}_{\zeta}(k) = \mathcal{A}_s \left(\frac{k}{k_{\rm cmb}}\right)^{n_s - 1}, \quad\quad \mathcal{P}^{(\rm v)}_h = r(k_{\rm cmb}) \mathcal{A}_s \left(\frac{k}{k_{\rm cmb}}\right)^{n_t},
\end{align}
where the CMB pivot scale is given by $k_{\rm cmb} = 0.05\, {\rm Mpc}^{-1}$. In the rest of this section, we will utilize \eqref{psvp} with the above parameter choices in order to explore the phenomenological implications of axion-gauge field dynamics during inflation. 

Moving the focus on to the sourced contributions to the power spectra, the location of the peak(s) depends on the time $N_*$ (the variable $N$ counts the number of e-folds before the end of inflation) at which the  axion's velocity becomes maximal. In particular, the scale that exits the horizon at this time $k_* = a(N_*) H(N_*)$  is as a free parameter that characterizes the scale dependence of the signal produced by the inverse decay of the vector fields. To fix $k_*$, we will assume that the axion is at its inflection point at $N_* = 31$ for the {\bf {\rm M1}} model. In the case of the {\bf {\rm M2}} model the axion will go through two inflection points corresponding to $N_* = 44$ and $N_* = 22$ and eventually settle to the minimum, long before the end of inflation. We will start our discussion from scalar perturbations.

\subsection{Enhanced curvature power spectrum and PBHs} 
The total scalar power spectrum in the model(s) we consider is given by 
\beq\label{Pzp}
\mathcal{P}_{\zeta}(k) = \mathcal{A}_s \left(\frac{k}{k_{\rm cmb}}\right)^{n_s - 1} \left[1 + \epsilon_{\phi}^2\, \mathcal{A}_s \left(\frac{k}{k_{\rm cmb}}\right)^{n_s - 1} \sum_i f^{(i)}_{2, \zeta}\left(\xi_{*},\frac{k}{k_{*}}, \delta\right)\right],
\eeq
{where again the summation inside the square parenthesis is optional for the ({\bf M2}) model, depending on whether the axion-like field probes multiple cliff-like regions before it settling down to its minimum.}
Following the procedures described in detail in Appendix \ref{AppA} and \ref{AppB} we compute the scale dependent contribution $f_{2,\zeta}$ to the power spectrum both in the WKB approximation and fully numerically. 
\begin{figure}[t!]
\begin{center}
\includegraphics[scale=0.75]{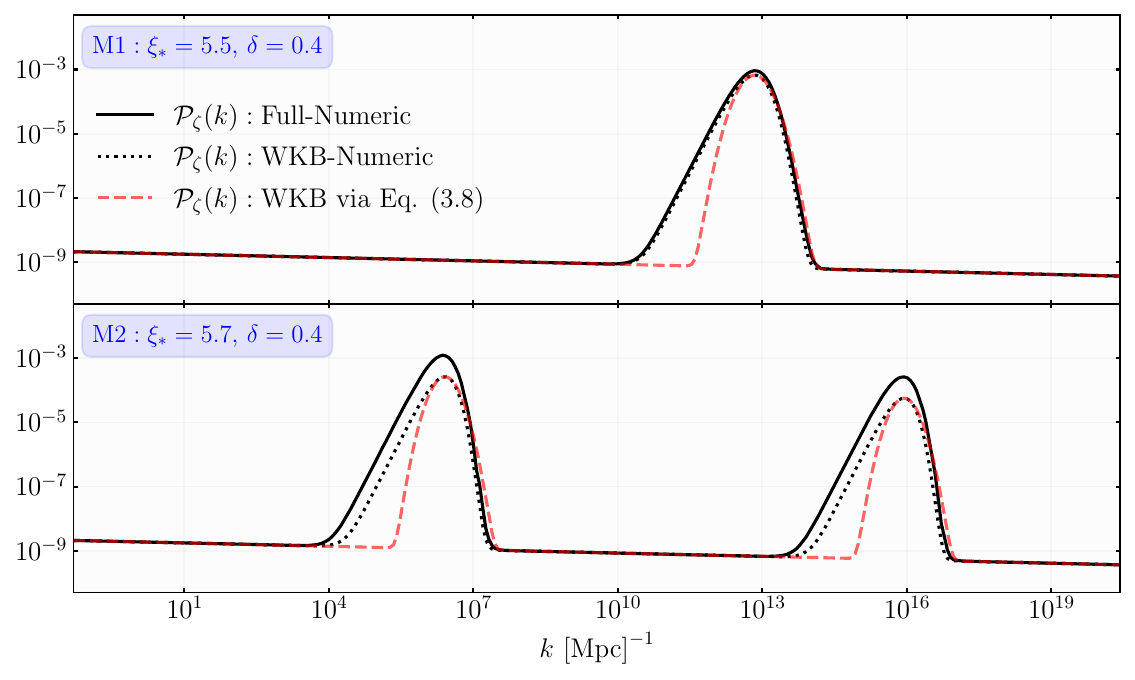}
\end{center}
\vspace*{-5mm}\caption{Total power spectrum of the curvature perturbation within three different approximations we are undertaking: Fully numeric (solid), numerically fitted normalization factor $N(\xi_*, x_*, \delta)$ (dotted) and the corresponding signal using the Gaussian fitting form \eqref{Nform} in the WKB approximation \eqref{Amsol} to gauge mode functions.\label{fig:Pzeta}}
\end{figure}

The profile of the total power spectrum for representative parameter choices ($\{\xi_*, \delta\}$)  is shown in Fig.~\ref{fig:Pzeta}. As can be anticipated from our discussion in the previous section, the total power spectrum computed using  log-normal fitting functions  for the vector amplitude \eqref{Nform} can accurately model the  spectrum computed via numerically fitted $N(\xi_*, x_*, \delta)$ only in the very close vicinity of the peak. In particular, when obtained via numerically fitted $N$, the sourced scalar power spectrum exhibits significantly larger power in the IR away from the peak and has a sharper drop in the UV as compared to result obtained using Eq.~\eqref{Nform}.

We also carry out fully numerical computations for the spectra  (see Appendix \ref{AppA} and \ref{AppB}). These are then used as  the benchmark with respect to which we evaluate the two  approximations presented above.
In the case of the scalar power spectrum of the M1 model, results obtained by employing the numerically fitted $N(\xi_*, x_*, \delta)$ within the WKB method are in very good agreement with the fully numerical calculation in terms of both the shape and peak amplitude. This is clear upon inspecting the top panel of Fig. \ref{fig:Pzeta}. In the case of the M2 model, the same approximation  matches (up to an $\mathcal{O}(1)$ factor) the overall amplitude of the benchmark result. In light of our discussion in section \ref{sec2p2}, these results are not surprising:  the WKB solution \eqref{Amsol} with \eqref{Eform} tends to underestimate the amplitude of the gauge modes (see Fig.\,\ref{fig:vmf}) in the M2 model, in particular in the time domain where the bulk of the enhancement takes place.

The inflationary models we study in this work, have been shown to produce PBH when the large fluctuations generated at small scales re-enter the horizon \cite{Garcia-Bellido:2016dkw, Ozsoy:2020ccy}. Needless to say, PBH production is extremely sensitive to the spectral shape of the scalar power spectrum \cite{Germani:2018jgr, DeLuca:2020ioi}. This highlights the importance of our result: a numerical approach to (at least) the normalization factor $N$ of the gauge field wave function is necessary to deliver an accurate result for the scalar (and tensor) power spectrum and the ensuing PBH production.
Given that the PBH distribution depends also on other important factors such as higher order statistic of the curvature perturbations  \cite{Garcia-Bellido:2017aan,Caravano:2022epk}, we leave a detailed calculations on the PBH abundance that takes into account our new results to future work.
\bigskip

Another interesting phenomenological consequence PBH forming inflationary models is the inevitable production of a stochastic GW background \cite{Ananda:2006af,Baumann:2007zm}. The enhanced spectrum of curvature fluctuations required to produce PBHs acts also as a GW source \cite{Kohri:2018awv}. The characterization of such GW background (see \eg \cite{Yuan:2019wwo, Cai:2019cdl, Ozsoy:2019lyy, Pi:2020otn}) along with the properties of its scalar sources (see e.g. \cite{Unal:2018yaa, Garcia-Bellido:2017aan}) is an important avenue for testing inflationary models such the ones we study in the present work. We do not pursue this path here as typically the scalar-induced GW signal is overshadowed by the GW production due to vector fields \cite{Garcia-Bellido:2017aan}. We refer the interested reader to \cite{Domenech:2021ztg} for a recent comprehensive review on scalar-induced GW backgrounds. In the next section, we focus on the GW background due to the excited gauge quanta of section \ref{sec3}.

\subsection{Chiral GW background from vector field sources}\label{sec4p2}

As mentioned at the beginning of this section, in the presence of particle production events that arise as a result of axion-gauge field dynamics, the total tensor power spectrum can be written as 
\beq\label{Phtot}
\mathcal{P}_h(k) = r(k_{\rm cmb})\,\mathcal{A}_s \left(\frac{k}{k_{\rm cmb}}\right)^{n_t} \left[1 + \frac{r(k_{\rm cmb})\,\mathcal{A}_s}{256} \left(\frac{k}{k_{\rm cmb}}\right)^{n_t} \sum_i \sum_{\lambda = \pm} f^{(i)}_{2, \lambda}\left(\xi_{*},\frac{k}{k_{*}}, \delta\right)\right],
\eeq
\begin{figure}[t!]
\begin{center}
\includegraphics[scale=0.7]{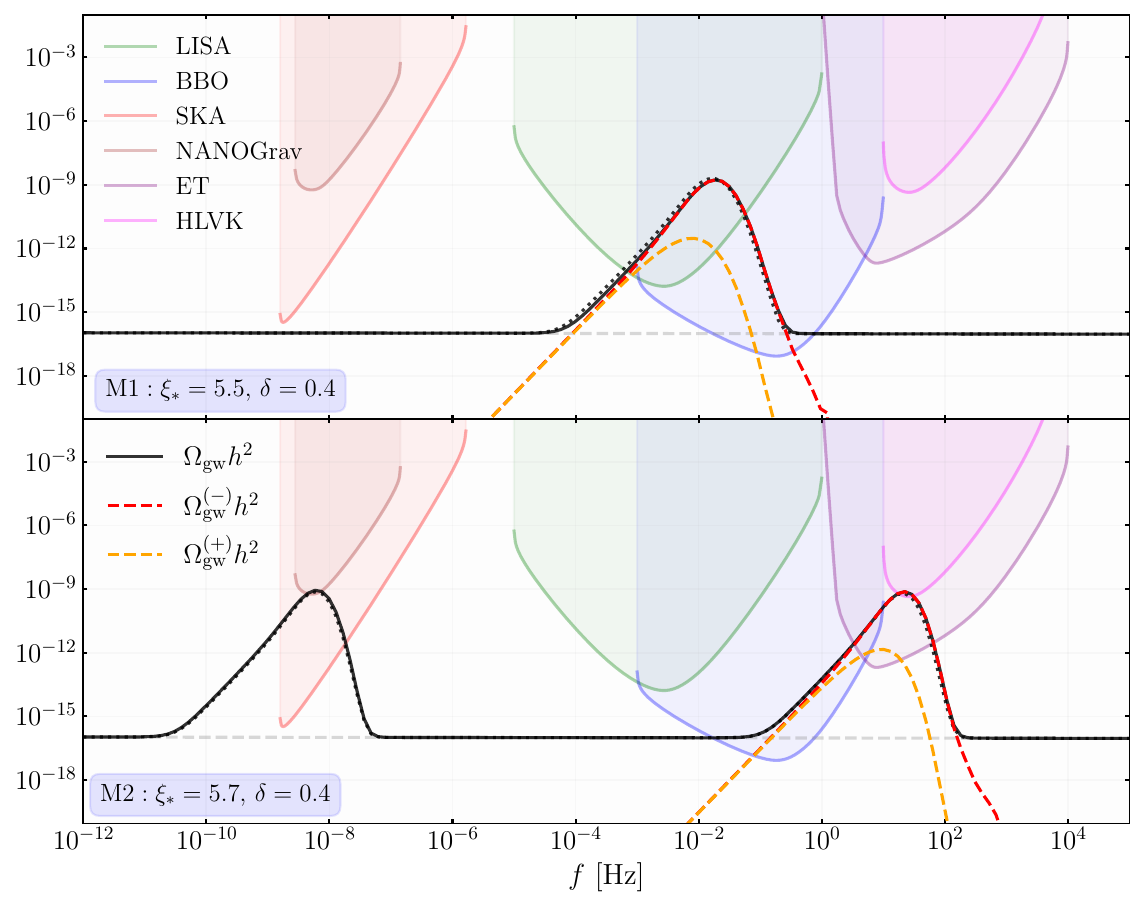}
\end{center}
\vspace*{-5mm}\caption{Total GW spectrum (solid black) together with its right (red dashed) and left (orange dashed) polarized \emph{sourced} components generated by axion-gauge field dynamics. The dotted curves corresponds to the GW spectrum calculated via the WKB approximated gauge mode functions. Power law integrated sensitivity curves \cite{Schmitz:2020syl} of various current and forthcoming GW probes are shown by shaded regions. The dashed gray line represents the vacuum GW background in the absence of particle production events. \label{fig:gwspec}}
\end{figure}

\noindent where, in full compliance with CMB constraints, we shall take $n_t = 3.75 \times 10^{-3}$, $\mathcal{A}_s \simeq 2.1 \times 10^{-9}$, $r(k_{\rm cmb}) = 0.03$ with $k_{\rm cmb} = 0.05\,{\rm Mpc}^{-1}$. Using Eq.~\eqref{Phtot} one may readily obtain
\beq\label{gwb}
\Omega_{\rm gw}(k)\, h^2 \equiv \left(\Omega_{\rm gw}^{(+)}(k) + \Omega_{\rm gw}^{(-)}(k)\right)h^2 = \frac{\Omega_{r,0}\,h^2}{24} \mathcal{P}_h(k),
\eeq
with $\Omega_{r,0} h^2 \simeq 4.2 \times 10^{-5}$ is the radiation density today\footnote{{Note that our expression assumes standard radiation domination for the post inflationary universe and we are neglecting changes to the effective number of degrees of freedom as they are irrelevant for $T\geq 100\;{\rm GeV}$ which corresponds to $f>4\cdot 10^{-4} \; {\rm Hz}$ and only mildly affect PTA scale GWs by a factor 2 enhancement.}}.  In Fig.~\ref{fig:gwspec} we present the GW background \eqref{gwb} in the frequency $f = k/2\pi$ domain relevant for current and forthcoming GW probes for the two scenarios we have been studying. In these plots, solid and dashed curves refers to the fully numeric approach whereas  dotted lines correspond to the GW signal obtained by adopting the WKB approximated gauge mode functions.

One can see from both panels in Fig. \ref{fig:gwspec} that the asymmetrical shape of the total GW spectrum around the peak of the signal is in accordance with the expectations outlined in section \eqref{sec2p2}. Gauge field fluctuations have a milder amplification for modes that exit the horizon before the axion reaches the inflection point(s) of its potential. This property  is inherited by the tensor perturbations for scales that precedes the peak(s) of the signal. Importantly, at such frequencies left and right polarization of the sourced GWs have a similar amplitude as clear from Fig. \ref{fig:gwspec}.

These results differ significantly from the GWs spectrum obtained using the Gaussian fitting form \eqref{Nform} in the gauge mode function \eqref{Amsol} (see Fig.~\ref{fig:f2s}) as is typically done in the literature (see \eg \cite{Namba:2015gja, Ozsoy:2020ccy, Ozsoy:2021onx, Ozsoy:2020kat, Campeti:2022acx}).
The very same plots in Fig. \ref{fig:gwspec} convincingly show that adopting a numerical fitting procedure for the normalization factor $N$  within the WKB approximation gives a GW template (dotted curves) that agrees very well with the signal obtained via the fully numerical approach (solid curves).
\begin{figure}[t!]
\begin{center}
\includegraphics[scale=0.65]{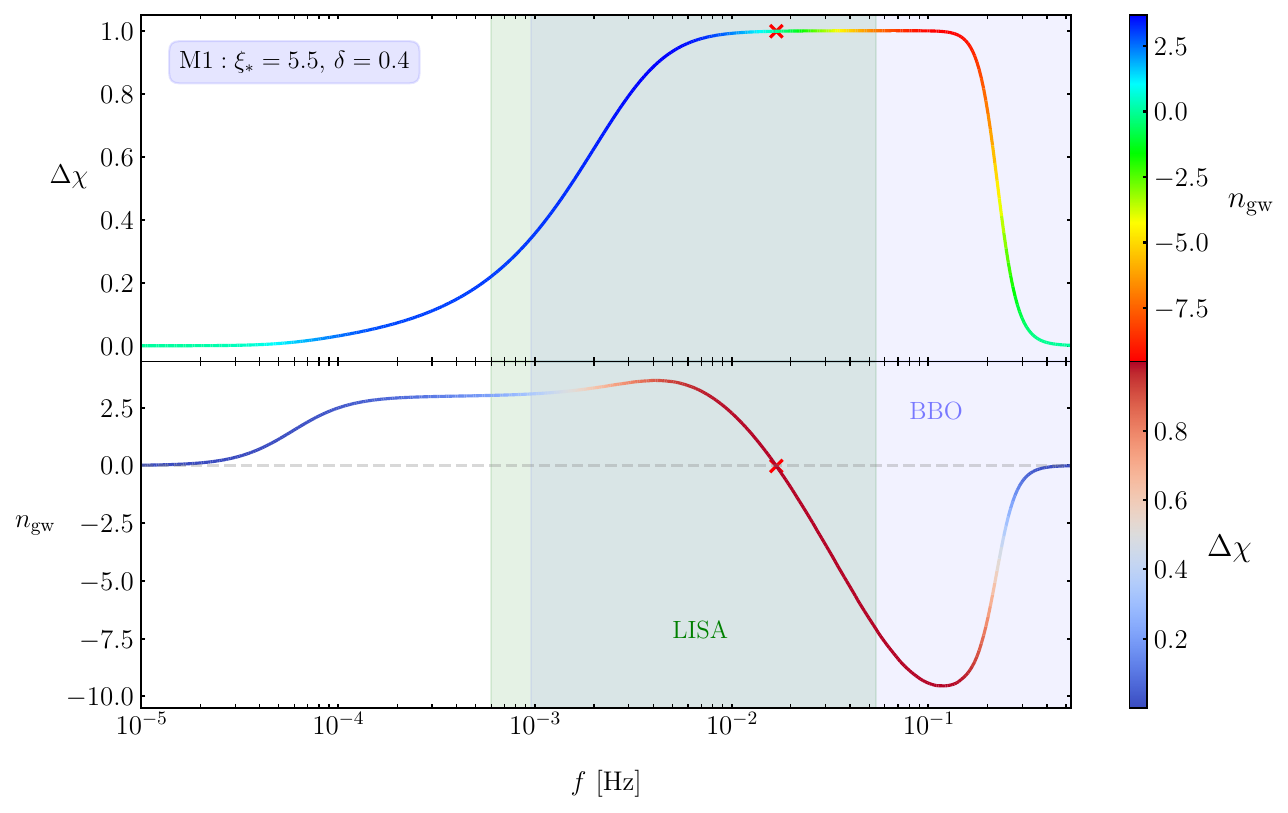}
\end{center}
\vspace*{-5mm}\caption{Chirality (top) and spectral index (bottom) of GW background produced in the {\bf M1} model of Fig. \ref{fig:gwspec}. For each plot, the value of $n_{\rm gw}$ (top) and $\Delta\chi$ (bottom) is shown along the curve using an appropriate color coding. Shaded bands indicates the range of frequency domain where the produced signal is within the power law integrated sensitivity curve of LISA (opaque  green) and BBO (opaque blue) probes. The red cross indicates the peak of corresponding GW signal where $n_{\rm gw} = 0$. \label{fig:dchiM1}}
\end{figure}

\medskip
\noindent{\bf The chirality of GWs around the peak.} The scale-dependence of the chirality of the GW signal is a very distinct, unique signature of the class of inflationary models we have been studying. To emphasize this point, we study the frequency dependence of the chirality for the two scenarios analysed in Fig.~ \ref{fig:gwspec}. The quantity we are interested in is the chirality parameter:

\beq\label{dchi}
\Delta \chi(f) = \frac{\mathcal{P}_{+}(f) - \mathcal{P}_{-}(f)}{ \mathcal{P}_{+} (f) +  \mathcal{P}_{-} (f)},
\eeq

\begin{figure}[t!]
\begin{center}
\includegraphics[scale=0.65]{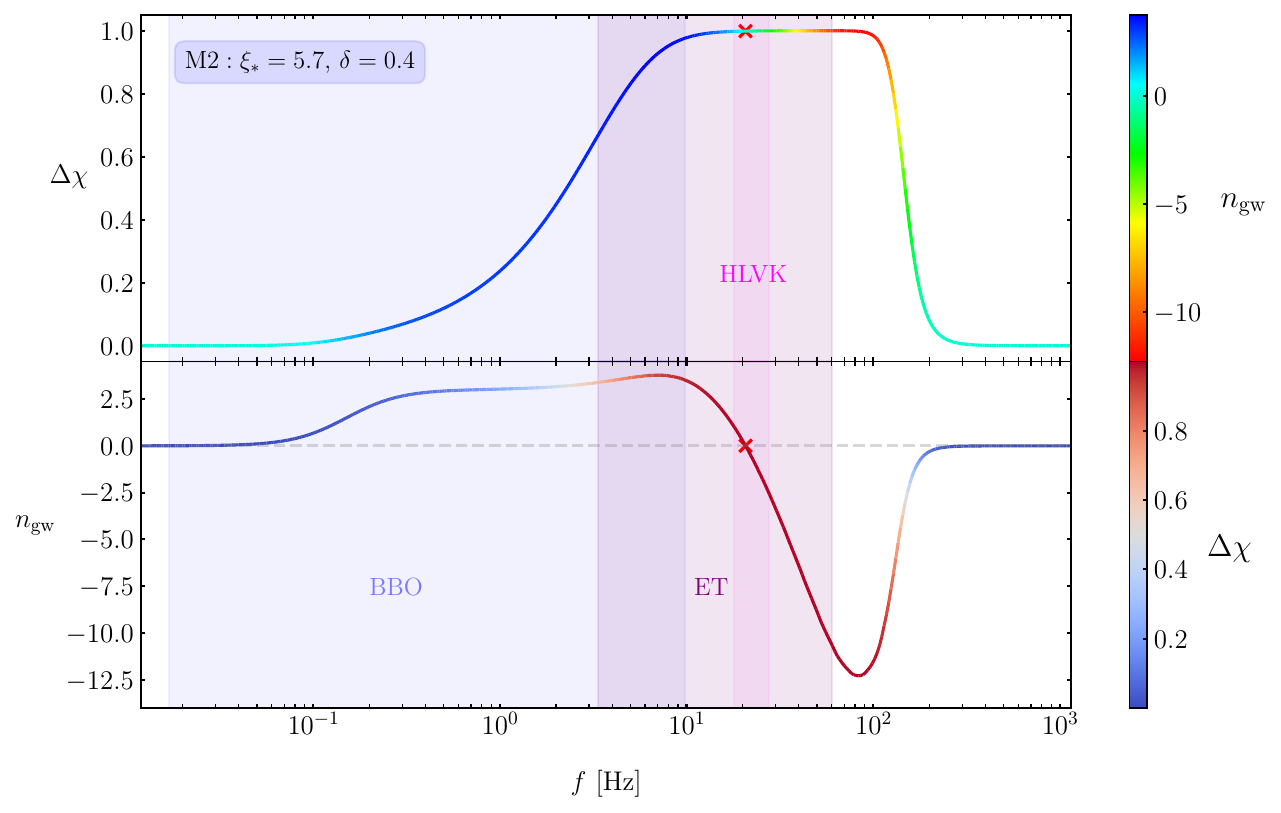}
\end{center}
\vspace*{-5mm}\caption{Chirality (top) and spectral index (bottom) of GW background produced in the second peak of {\bf M2} model (see Fig. \ref{fig:gwspec}). For each curve, the value of $n_{\rm gw}$ (top) and $\delta\chi$ (bottom) is shown along the curve using an appropriate color coding. Shaded bands indicate the range of frequency domain where the signal is within the power law integrated sensitivity curve of BBO (opaque  blue), ET (opaque purple) and LIGO-Virgo-Kagra collaboration (opaque magenta) missions.\label{fig:dchiM2}}
\end{figure}

\noindent where we switched to frequency domain using $k = 2\pi f$ and where $\mathcal{P}_{\lambda}$ is the total tensor power spectrum \eqref{tps} for each GW polarization. The spectral index of the GW background is another crucial quantity. Our findings as well as previous work in \cite{Garcia-Bellido:2023ser} suggest the usefulness of exploring the relation between the spectral index and the scale-dependent chirality.

One can schematically see from Fig.~\ref{fig:gwspec} that, as long as the sourced contribution dominates over the vacuum, the more blue-tilted is the spectrum, the smaller the chirality will be. One is especially interested in the frequency dependence of the spectral index around the peak of the signals (see Fig. \ref{fig:gwspec}) relevant for current and upcoming GW probes. We characterize it in the frequency domain as
\beq\label{si}
\frac{\d \ln \Omega_{\rm gw}}{\d \ln f} \equiv n_{\rm gw}(f).
\eeq
A study of the correlation between the chirality \eqref{dchi} and the corresponding spectral index \eqref{si} could  provide unique information about the relevant field content during inflation and help us decouple the primordial signal from possible \cite{Christensen:2018iqi} and perhaps inevitable \cite{Chen:2018rzo,NANOGrav:2023hvm} astrophysical sources.

In Figs. \ref{fig:dchiM1} and \ref{fig:dchiM2}, we present the frequency dependence of the chirality parameter \eqref{dchi} and spectral index $n_{\rm gw}$ for the single {\bf M1} and double peak {\bf M2} scenarios. In the {\bf M2} case,  we focus our attention only on the second peak because it is the relevant one for frequencies associated with BBO \cite{Crowder:2005nr} and Einstein Telescope \cite{Hild:2010id}. To emphasize the relation between the chirality on the spectral index, we use a color coding that changes depending on the value of each quantity. {It is clearly visible from the plots that under a blue-tilted GW spectrum the chirality $\Delta \chi$ quickly rises until the spectral tilt becomes zero (highlighted with a red cross in Figs.  \ref{fig:dchiM1} \& \ref{fig:dchiM2}), and then continues to rise at a slower pace reaching values closer and closer to unity for negative values of the spectral index. In the far infrared and UV where the spectra has very mild negative slopes, the net chirality of the GW signal goes to zero. This is expected as those regions of the spectrum are controlled by the vacuum fluctuations of the metric which are not parity violating. This dynamical behaviour of the chirality and spectral index can occur at frequencies where the GW background amplitude  exceeds the sensitivity of multiple GW probes, as is indicated by the colored shaded regions in Figs.~\ref{fig:dchiM1} and \ref{fig:dchiM2}.} Fortunately, there are several ways to test GW chirality at interferometer scales. One can, for example, use the effective non-planar geometry resulting from correlating interferometers at a different geographic locations \cite{Smith:2016jqs}. Even in the case of a single planar interferometer, the  kinetically induced dipolar anisotropy due to the solar system motion with respect to the cosmic rest frame confers a certain sensitivity to circular polarization \cite{Domcke:2019zls}.

The chirality-\textit{vs}-GW spectral tilt relation is particularly relevant for the class of models under scrutiny in this work. The non-linear nature of the GW sourcing via gauge fields  makes the chirality scale-dependent even in the regime where the sourced contribution is clearly dominant over the vacuum. This is a unique, striking, signature associated with the presence of Abelian gauge fields. It is well-known that a fully chiral $\chi=1$ GW signal is strongly suggestive of a Chern-Simons coupling during inflation. One key conclusion that can be drawn from our results is that the detection of a frequency-dependent chirality\footnote{In the frequency range where the sourced contribution dominates.} gives us precious additional information: it points to the Abelian nature of the gauge group. Conversely, detecting a completely non-chiral signal over a small frequency range should not automatically lead one to exclude the CS possibility. It is in this context that having a map between chirality and GW spectral tilt
can be particularly useful. The patterns seen in Figs.~\ref{fig:dchiM1} \& \ref{fig:dchiM2} point to where, in terms of the values of the spectral tilt, one should expect chirality to emerge and reach its maximum. It is then very useful to derive a quantitative (analytical) relation between $n_{\rm gw}$ and GW chirality $\Delta \chi$. This is what we turn to do next.

\begin{figure}[t!]
\begin{center}
\includegraphics[scale=0.65]{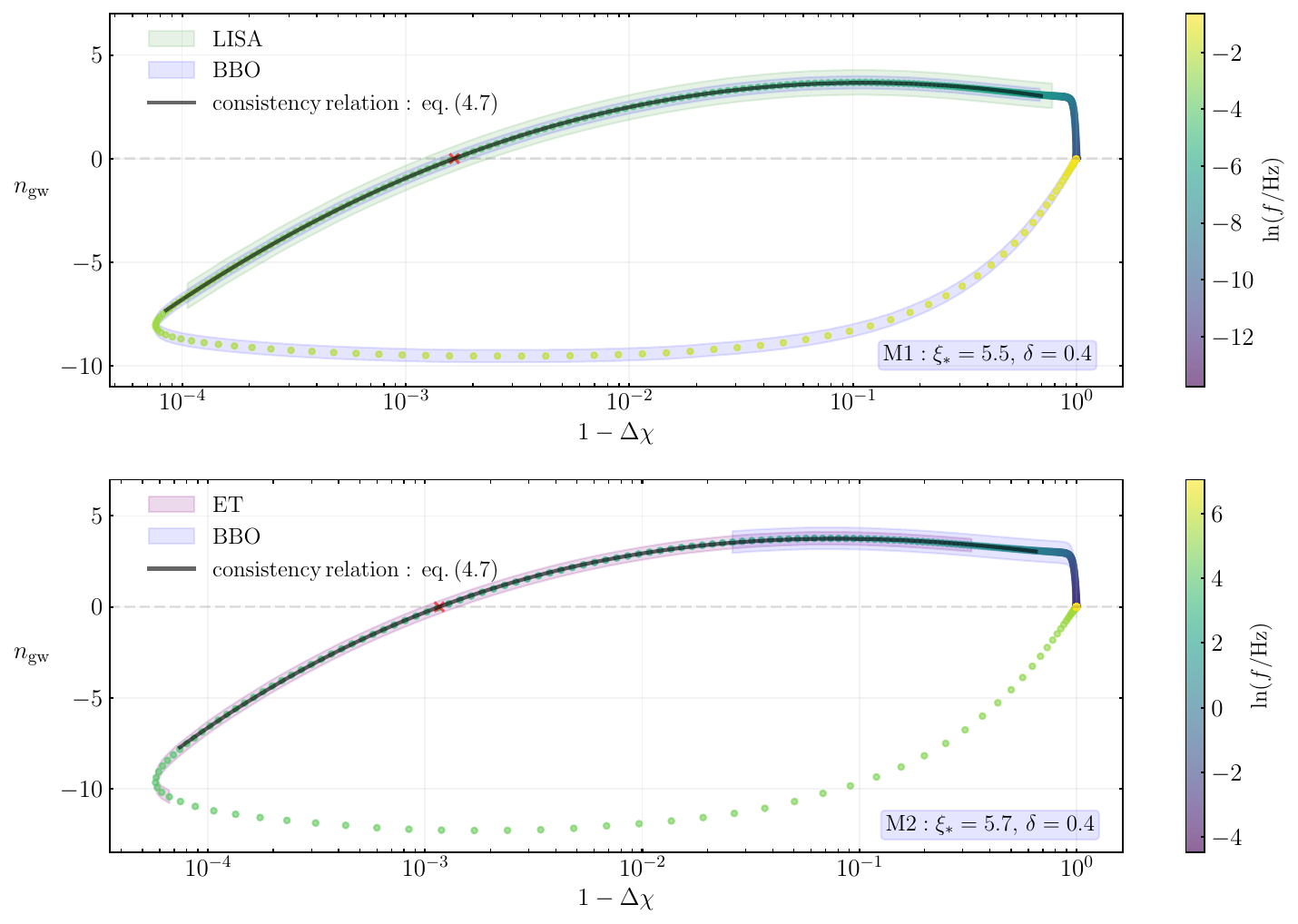}
\end{center}
\vspace*{-5mm}\caption{{Evolution of scale GW signal from IR to UV scales (using log-frequency scale) in the $1-\Delta\chi$ vs $n_{\rm gw}$ plane. Consistency condition \eqref{ngwform} is shown by solid black curve, parametrizing the evolution of the spectral tilt $n_{\rm gw}$ in terms of the excess chirality as the GW signal rise and falls around its peak value (shown by the red cross) where $n_{\rm gw} = 0$. Overlapping bands are shown to indicate the frequency range where the signal is above the sensitivity curves of forthcoming GW probes we consider in Figure \ref{fig:gwspec}.}  \label{fig:dchivsngw}}
\end{figure}

\noindent{\bf A consistency relation.}  A relation between chirality and spectral tilt would help break the degeneracy between various astrophysical and primordial sources and provide compelling evidence for GWs generated by axion-(Abelian) gauge field dynamics. We find it convenient derive such \emph{consistency relation} around the peak of the sourced GW signal for both models considered in this work. Specifically, during the rise and fall of the sourced spectra (e.g. disregarding the vacuum contribution), $3\lesssim n_{\rm gw} \lesssim -7.3$, we found that a third order polynomial accurately describes the map between the spectral index and the logarithm of the \emph{excess} chirality $\ln(1-\Delta\chi)$:
\begin{align}\label{ngwform}
n_{\rm gw} = \sum_{p = 0}^{3} c_{p}\, [\, \ln(1-\Delta\chi)\, ]^{p},
\end{align}
with we provide $c_p$ coefficients in Table \ref{tab:ngwtab}. Notice that the sign and absolute value of the fitting coefficients are very close to each other implying a universal behaviour shared by two models for a chirality range  $0.3 \lesssim \Delta \chi \lesssim 1$. {We illustrate these points in Fig. \ref{fig:dchivsngw} where the evolution of the GW signal in the $1-\Delta\chi$ vs $n_{\rm gw}$ plane is shown for both scenarios we consider in Fig. \ref{fig:gwspec}. We observe that far in the IR, chirality is vanishing ($\Delta\chi \sim 0$ or $1-\Delta\chi \sim 1$) while $n_{\rm gw} \lesssim 0$. As the vector field sources become active, chirality $\Delta\chi$ of the GW signal grows along the positive $n_{\rm gw}$ branch while maintaining its growth for frequencies following the peak where $n_{\rm gw} < 0$. Eventually particle production shuts down and hence the GW signal returns to the origin of the ``phase space"  where $\Delta\chi \to 0$ ($1-\Delta \chi \to 1$) and $n_{\rm gw}\lesssim 0$. The accuracy of the consistency relation \eqref{ngwform} around the maximum of the GW signal ($n_{\rm gw} = 0$)  -- shown by the solid dark curve -- is visible. Furthermore, the potential coverage of the signal with multiple probes indicated by the colored thick bands together with its parametrization by the consistency relation we derived in this work is particularly encouraging.}

The attentive reader may wonder about the universality of Eq.~(\ref{ngwform}) for different choices of the parameter $\delta$, which essentially parametrizes the width of the GW signal (and the mass of the axion around its global minimum) in the vicinity of the peak. We expect this relation to hold, up to order one changes in the $c_p$ coefficients, for different values of $\delta$ so long as we remain in the  $\delta \ll3$ regime, where we are nevertheless confined in light of the requirement that the axion be slowly rolling. Naturally, a lower bound on $\delta$  emerges from requiring  $\xi$ not be constant (this corresponds to $m_{\chi}\lesssim H$), which would mean a fully chiral signal as per the original analysis of \cite{Sorbo:2011rz}. We expect our consistency relation to be especially  useful when used within these two qualitative thresholds.

\begin{table}[t!]
\begin{center}
\begin{tabular}{| c | c | c | c | c |}
\hline
\cellcolor[gray]{0.9}&\cellcolor[gray]{0.9}$c_0 \simeq$&\cellcolor[gray]{0.9}$c_1 \simeq$&\cellcolor[gray]{0.9}$c_2 \simeq$&\cellcolor[gray]{0.9}$c_3 \simeq$\\
\hline
$\cellcolor[gray]{0.9} \textrm{M1} $ & \scalebox{0.95}{2.76}&\scalebox{0.95}{-0.82}&\scalebox{0.96}{-0.179}&\scalebox{0.96}{0.0025}\\\hline 
$\cellcolor[gray]{0.9}\textrm{M2}$ & \scalebox{0.95}{2.76}&\scalebox{0.95}{-0.71}&\scalebox{0.95}{-0.104}&\scalebox{0.95}{0.0092}\\\hline 
\hline
\end{tabular}
\caption{Coefficients of the polynomial fitting form \eqref{ngwform} that relates the spectral index $n_{\rm gw}$ to the chirality $\Delta \chi$ of GW signal produced by the axion-gauge field dynamics. \label{tab:ngwtab}}
\end{center}			
\end{table}

\noindent{\bf The validity of the small back-reaction regime.} Our derivations for the scale
dependent templates of the total power spectra (and their properties) assume that the back-reaction of vector field fluctuations on the background evolution of the spectator axion is negligible. On the other hand, as can be seen from Fig.~\ref{fig:norms}, the gauge field amplification needs to be sufficiently strong to leave observable effects at small scales (see Figs. \ref{fig:Pzeta} and \ref{fig:gwspec}). It is therefore necessary to have a well-defined criterion to justify our small back-reaction assumption. Identifying such criterion is important also in light of the highly non-trivial dynamics that has been described in the strong back-reaction regime of axion inflation \cite{Cheng:2015oqa,Notari:2016npn,Sobol:2019xls,DallAgata:2019yrr,Domcke:2020zez,Caravano:2021bfn,Gorbar:2021rlt,Durrer:2023rhc,Caravano:2022epk,Garcia-Bellido:2023ser,vonEckardstein:2023gwk,Domcke:2023tnn}. Furthermore, lattice simulations \cite{Figueroa:2023oxc} have shown that in the strong back-reaction regime the fields spatial gradients can become large and have a dramatic impact on the background fields evolution\footnote{This is certainly the case at least for a certain range of values of the coupling constant $\lambda$.}.

The results above have largely been derived for the case of an axion-inflation. However, it is reasonable to assume that analogous conclusions apply for the spectator case we study in the present work. The mode-by-mode approach we employed  disregards the spatial dependence of the axion, and therefore we limit our analysis in the ``safe'' small back-reaction regime for which the spatial gradient of the axion do not play an important role. In such a regime, the mode-by-mode treatment captures the full dynamics of the system.

Another reason we focus on the small back-reaction regime is that we want to stress the fact that the physics underpinning the scale-dependent chirality and the asymmetrical shape of the  power spectra is in no way associated with the non-trivial dynamics observed in the strong back-reaction regime. Instead, the physics we study here represents a refinement of the well studied axion-U(1) dynamics during inflation in the small back-reaction regime. We explicitly take into account these back-reaction effects to show that they can indeed be neglected for the parameter choices we consider in this work. In what follows, we will closely follow \cite{Anber:2009ua,Peloso:2016gqs,Garcia-Bellido:2023ser}, and refer the reader to the same works for additional details. 

In the mean field approximation, back-reaction effects due to particle production can be represented by the expectation value of the dot product of electric and magnetic components of the vector field as  
\begin{equation}
    \ddot{\chi}+3H \dot{\chi}+U_\chi'(\chi)=\frac{\lambda}{f}\left\langle\vec{E}\cdot\vec{B}\right\rangle\, ,
\end{equation}
where
\begin{equation}
\left\langle\vec{E}\cdot\vec{B}\right\rangle=-\frac{1}{4\pi^2a^4}\int {\rm d}k\, k^3\frac{\partial}{\partial \tau}\big|A_{-}(\tau,k)\big|^2.
\end{equation}
Since the dominant part of the gauge mode production is localized around $x_* = k/k_*$, corresponding to times when the axion rolls through the inflection points of its potential, we can derive a ``normalized spectral back-reaction'' ${\cal B}$ by integrating over these modes as 
\begin{equation}
    \frac{\lambda\left\langle\vec{E}\cdot\vec{B}\right\rangle}{f \,U'_\chi(\chi)}\equiv\int {\rm d}\ln{x_*}\;{\cal B}(\tau/\tau_*,x_*)\;.
    \label{eq:backreaction}
\end{equation}
\begin{figure}[t!]
\begin{center}
\includegraphics[scale=0.59]{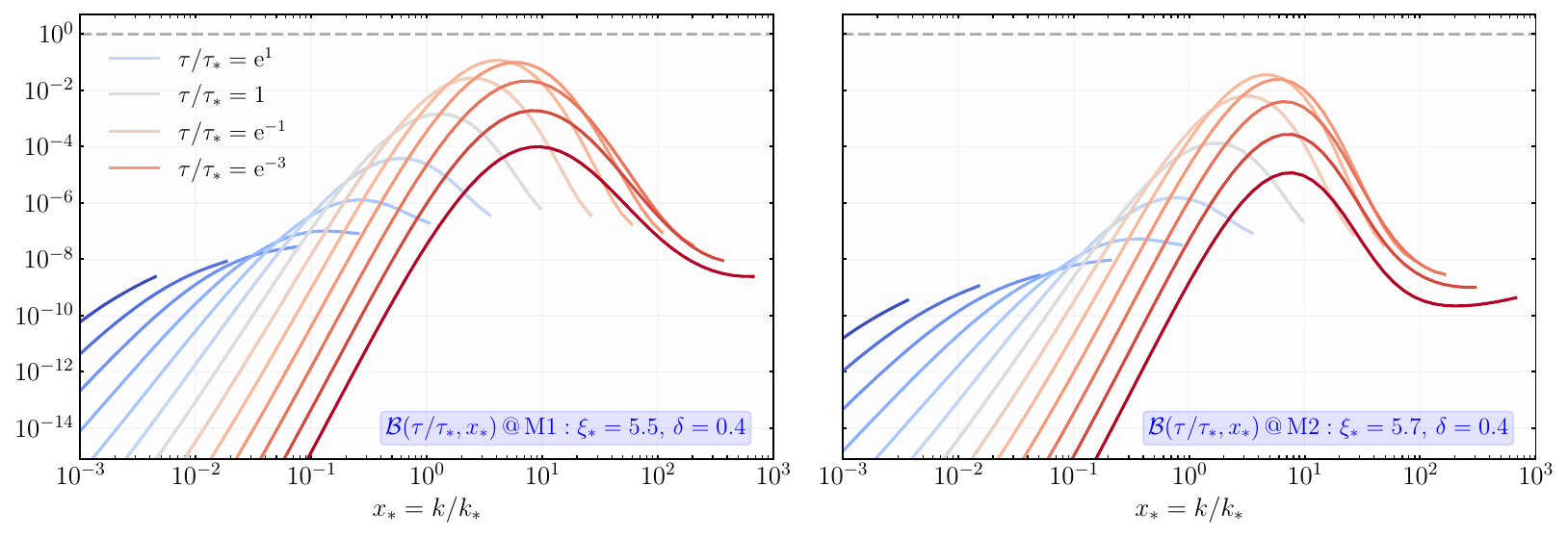}
\end{center}
\vspace*{-5mm}\caption{Normalized spectral back-reaction $\mathcal{B}$  \eqref{eq:backreaction} evaluated at 13 equally spaced moments of time symmetrically distributed around the time $\tau_*$, \ie when the axion traverses the inflection point(s).\label{fig:B}}
\end{figure}
This normalization implies that when the integral of ${\cal B}$ in $\ln x_*$ space is close to one, the back-reaction  significantly affects the equation of motion of the axion. We require that the right hand side of (\ref{eq:backreaction}) is at least an order of magnitude smaller than unity at every moment $\tau/\tau_*$ in time during the roll of the axion around the inflection point(s). 

We numerically compute $\mathcal{B}$ for parameter choices we adopt in Figs. \ref{fig:Pzeta}-\ref{fig:dchiM2}. For both models we consider, the resulting normalized spectral back-reaction for various moments in time is shown in Fig.~\ref{fig:B}. The upper limit (in $x_* = k/k_*$ space) of the normalized spectral back-reaction curves in the figures are given by the threshold between stable and unstable modes $k_{\rm thr}\equiv - a \lambda \dot{\phi}/f \Rightarrow x^{\rm thr}_{*} = 2\xi(\tau) / (\tau/\tau_*)$. The threshold is increasing over time (as $\tau/\tau_* \to 0 $), hence the different upper limits to the back-reaction integral in $\ln x_*-{\rm space}$ at different times. We have plotted the normalized spectral back-reaction for 13 different moments in time in 1 e-folding increments symmetrically distributed around $\tau=\tau_*$. One notable feature is that there is a delay effect between the moment of largest particle production parameter $\xi=\xi_*$ and greatest back-reaction. Additionally, the maximally enhanced mode is shifted (towards the UV) with respect to the mode ($ x_* = k/k_*  = 1$) that exits the horizon exactly when effective coupling parameter peaks, $\xi(\tau_*)=\xi_*$. This is another manifestation of the delay effect studied in depth in \cite{Domcke:2020zez}.

 The small value of the logarithmic integral in \eqref{eq:backreaction} ensures the validity of the phenomenology we presented in section \ref{sec4},  and justifies our approach of neglecting the back-reaction of the amplified gauge modes. Extending the results presented above to the strong back-reaction regime for the models above is an interesting direction which we are planning on exploring in future work.

\section{Discussion and Conclusions}
\label{sec:con}
The presence of additional field content in the form of an axion-gauge sector during inflation can lead to an array of interesting phenomenology across a wide range of scales relevant for cosmological observations, including the production of primordial black holes and chiral GW background.  This phenomenology is particularly sensitive to the details of the particle production processes in the gauge sector.

In this work we explored in detail the pitfalls related to the use of an important approximation  commonly found in the literature on vector field production by rolling axions. 
We went beyond the standard parametrization of the amplitude of vector modes via the log-normal form \eqref{Nform}; we did so by \emph{first} computing the scale dependent amplitude $N(k/k_*)$ using a numerical fitting procedure within the WKB approximation \eqref{Amsol}. 
We refer to this method as semi-analytic hybrid method (or {\rm WKB-numeric}). As elaborated in depth in section \ref{sec3}, the inability of the standard approach to accurately reproduce the findings of the hybrid (and the fully numerical) method is especially evident from the frequency dependence of the sourced scalar and tensor spectra  (see Fig. \ref{fig:f2s}). We found that when sourced spectra dominate over their vacuum counterparts, 2-point scalar and tensor functions exhibit a milder slope in the IR region preceding the peak w.r.t. the UV regions. This key feature is not captured by the standard approximation.  For the sourced tensor spectrum this IR/UV asymmetry has an important consequence: a net suppression of the chirality of the total GW spectrum in the IR (see also \cite{Garcia-Bellido:2023ser}). 

Going Beyond the WKB solutions of Eq.~\eqref{Amsol}, we computed fully numerically the gauge field production and the resulting power spectra of tensor and scalar perturbations (see Appendix \ref{AppA} and \ref{AppB}) in  both models presented in section \ref{sec2p1}. Our findings in sections \ref{sec3} and \ref{sec4} confirm the validity of the hybrid approach with mild caveats on the scalar power spectrum of the M2 model (see Figs. \ref{fig:Pzeta} and \ref{fig:gwspec}). 
The improved templates we derived for primordial  perturbations can affect the physics of primordial black holes (\eg induced GWs, PBH abundance etc.) and the primordial GW background generated by hidden axion-gauge sectors. In this work, we chose to focus on the latter and study in detail \emph{chirality} and \emph{spectral index} of the tensor modes at scales relevant for current and upcoming GW interferometers. In order to aid future experimental efforts to detect primordial GWs from axion-gauge field models, we derived a \emph{consistency relation} that ties the tensor spectral index to the expected GW chirality. We expect this relation to be (qualitatively) universal as long as the axion remains light during inflation, $m_\chi \lesssim H$.  

Our findings can be expanded in several directions. First and foremost, it would be interesting to revisit earlier studies on sourced GWs at CMB scales (see \eg \cite{Namba:2015gja, Ozsoy:2020ccy, Campeti:2022acx}). The expected $\langle BB\rangle$ correlation, for example, might as well be sensitive to the IR/UV asymmetry we have pointed out in the gauge field mode functions. For the very same reasons, it will be important to tackle  $\langle EB\rangle$ and $\langle BT\rangle$ correlations: these are precisely those sensitive to chirality and it is paramount to explore the effect that a scale dependent-chirality would have on such observables. Implementing the precise momenta dependence of the gauge fields wave-function is bound to be crucial also in determining strong back-reaction effects. This is certainly a direction we plan to pursue in future work. Another realm where our results may have repercussions is that of perturbativity bounds (see \eg \cite{Peloso:2016gqs, Campeti:2022acx}): these limit the region of the model parameter space accessible without resorting to non-perturbative techniques. In particular, perturbativity of higher order corrections to the gauge field propagator can deliver the most restrictive of all such constraints \cite{Ferreira:2015omg}, which underscore the importance of accurately modelling the wave-number dependence of gauge modes.

In this work we focused on two-point function and so one might wonder how would our findings reflect on higher-order correlators. This question is relevant already at the level of currently operating experimental probes as existing bounds on GW non-Gaussianities from the Planck mission have placed  constraints on axion gauge field models \cite{Agrawal:2018mrg}. Refining purely GW \cite{Shiraishi:2016yun} as well as mixed non-Gaussianity \cite{Ozsoy:2021onx} calculations by means of our semi-analytical or fully numerical approach will be certainly worthwhile. As repeatedly mention in the text, another natural arena to revisit is that of PBH production and the corresponding scalar-induced GW spectrum \cite{Garcia-Bellido:2016dkw, Garcia-Bellido:2017aan, Ozsoy:2020kat, Ozsoy:2023ryl}. We leave this to upcoming work.

\acknowledgments
 O\"O is  supported by the “Juan de la Cierva” fellowship IJC2020-045803-I. MF and AP acknowledge support from the “Consolidaci\'{o}n Investigadora” grant CNS2022-135590. MF acknowledges support also from the “Ram\'{o}n y Cajal” grant RYC2021-033786-I. This work is partially supported by the Spanish Research Agency (Agencia Estatal de Investigaci\'{o}n) through the Grant IFT Centro de Excelencia Severo Ochoa No CEX2020-001007-S, funded by MCIN/AEI/10.13039/501100011033. 
\newpage
\begin{appendix}
\section{Numerical recipes for vector field production}\label{AppA}
In this appendix, we provide details for the numerical methods we utilize to determine the normalization function $N(\xi_*, x_*, \delta)$ within the WKB approximation \eqref{Amsol}, as well as the vector mode evolution through a fully numerical approach. Focusing on the polarization amplified by the rolling axion, we start off by a redefinition of mode functions:
\beq\label{rd}
\bar{A}_{-}(x) \equiv \tilde{A}_{-}(x)\, {\rm e}^{-ix},\quad\quad \tilde{A}_{-}(x) \equiv \sqrt{2k}\, A_{-}(\tau,k).
\eeq
In terms of the barred variable, the equation of motion \eqref{meqgf} for the complex vector field can be written as
\beq\label{Abarc}
\frac{\d \bar{A}_{-}}{\d x^2} + 2i \frac{\d \bar{A}_{-}}{\d x} - \frac{2\xi(x)}{x} \bar{A}_{-} = 0,
\eeq
where the effective coupling $\xi$ between gauge and axion-like field is given as \eqref{xip}. In terms of the new variable, the equation can be numerically solved using the corresponding Bunch Davies initial conditions: 
\beq\label{Abarcin}
\bar{A}_{-}(x_{\rm in}) = 1,\quad\quad \bar{A}'_{-}(x_{\rm in}) = 0,
\eeq
where prime indicates a derivative with respect to the argument. WKB approximated mode functions \eqref{Amsol} satisfy an important relation that we can utilize to determine the normalization factor using the numerical solutions to the barred variable. This relation read as 
\beq\label{ir}
\frac{\d A_{-}^{\prime}(\tau, k)}{\d \tau}=\sqrt{\frac{2 k\, \xi(\tau)}{-\tau}} A_{-}^*(\tau,k).
\eeq
Using \eqref{ir} and \eqref{rd} with \eqref{Amsol}, the normalization factor can be written as 
\beq\label{Nwkb}
N(\xi_*, x_*, \delta) = \sqrt{\big|\bar{A}_{-}(x)(\bar{A}'_{-}(x)+ i \bar{A}_{-}(x))\big|}\, \exp\left[{2E(x)\sqrt{2\xi_* x_*}}\right].
\eeq
Since the WKB form \eqref{Amsol} characterize the late time solution of the mode functions away from the turning point $x \ll x_*$ by definition, the normalization factor in \eqref{Nwkb} can be obtained numerically by solving \eqref{Abarc} with \eqref{Abarcin} for a given $\xi_*$ and $\delta$ for a range of $x_* = k/k_*$ values and matching this solution to the expression \eqref{Nwkb} at late times $x \to 0$. The resulting scale dependence of the normalization factor for representative parameter choices is shown by the solid curves in Fig. \ref{fig:norms}.

\medskip
\noindent{\bf Fully numeric evaluation of gauge fields.} In order to check the validity of the WKB solution \eqref{Nwkb}, we also implement a fully numerical method to compute the gauge field evolution which we in turn to utilize the evaluate the resulting 1-loop scalar and tensor power spectra (see Appendix \ref{AppB}). For this purpose we find it convenient to split the equation \eqref{Abarc} into its real and imaginary parts which we solve together using the following coupled set of equations: 
\begin{align}\label{Abarcn}
\nn \fr{\d \bar{A}^{(r)}_{-}}{\d x^2} &- 2 \frac{\d \bar{A}^{(i)}_{-}}{\d x} - \frac{2\xi(x)}{x} \bar{A}^{(r)}_{-} = 0,\\
\frac{\d \bar{A}^{(i)}_{-}}{\d x^2} &+ 2 \fr{\d \bar{A}^{(r)}_{-}}{\d x} - \frac{2\xi(x)}{x} \bar{A}^{(i)}_{-} = 0,
\end{align}
where $\{r,i\}$ stands for the real and imaginary part of the barred variable $\bar{A} = \bar{A}^{(r)} + i \bar{A}^{(i)}$. The set of equations can be solved numerically by initializing the modes in the Bunch Davies vacuum deep inside the horizon as 
\beq\label{Abarcinn}
\bar{A}^{(r)}_{-}(x_{\rm in}) = 1,\quad\quad \bar{A}^{(i)}_{-}(x_{\rm in}) = 0, \quad\quad \frac{\d \bar{A}^{(r/i)}_{-}(x)}{\d x} \Bigg|_{x_{\rm in}} = 0.
\eeq
We note that due to $\xi_*, x_*$ and $\delta$ dependence of the effective coupling $\xi$ \eqref{xip} in \eqref{Abarcn}, the numerical solutions inherently have dependence of these parameters that characterize the roll of the axion field. To derive the phenomenological results we present in section \ref{sec4}, we adopt $\delta = 0.4$ and evolve \eqref{Abarcn} with the initial conditions \eqref{Abarcinn} for a range of scales $x_* = k/k_*$ focusing on $3.5\leq \xi_*\leq 6.5$, in order to accurately capture the time evolution of the amplified gauge field modes. Keeping this procedure in mind, we will omit the dependence of the vector field on $\{\xi_*, x_*, \delta\}$ in what follows. 

In the rest of this Appendix, we would like to note electric and magnetic fields of the dark vector fields which appear as sources in the cosmological correlators we study in the next section. For this we define the electromagnetic fields as 
\beq
\hat{E}_i(\tau, \vbf{x}) = - a^{-2}\, \partial_\tau\hat{A}_i(\tau, \vbf{x})\quad\quad \hat{B}_i(\tau, \vbf{x}) =  a^{-2}\,\epsilon_{ijk} \partial_j\hat{A}_k(\tau, \vbf{x}),
\eeq
and write their Fourier space counterparts as 
\begin{align}\label{EB}
\nn & \hat{E}_i(\tau, \vbf{k})=\sqrt{\frac{k}{2}} \frac{\epsilon_i^{(-)}(\vbf{k})}{a(\tau)^2} \left[\tilde{A}'_{-}(-k\tau)\, \hat{a}_{-}(\vbf{k}) + \tilde{A}'_{-}(-k\tau)^{*}\, \hat{a}^{\dagger}_{-}(-\vbf{k}) \right], \\
& \hat{B}_i(\tau, \vbf{k})=-\sqrt{\frac{k}{2}} \frac{\epsilon_i^{(-)}(\vbf{k})}{a(\tau)^2} \left[\tilde{A}_{-}(-k\tau)\, \hat{a}_{-}(\vbf{k}) + \tilde{A}_{-}(-k\tau)^{*}\, \hat{a}^{\dagger}_{-}(-\vbf{k}) \right],
\end{align}
where $\tilde{A}_{-}(x) = \tilde{A}_{-}(-k\tau)$ is defined as in \eqref{rd} and primes denote derivatives with respect to the arguments of the mode functions. For the convenience of notation we define the following operators that will be instrumental for the computations we present in the next section:
\begin{align}\label{OEB}
\nn &\hat{\mathcal{O}}_E(\tau,\vec{q}) \equiv \left[\tilde{A}'_{-}(-q\tau)\, \hat{a}_{-}(\vbf{q}) + \tilde{A}'_{-}(-q\tau)^{*}\, \hat{a}^{\dagger}_{-}(-\vbf{q}) \right],\\
&\hat{\mathcal{O}}_B(\tau,\vec{q}) \equiv \left[\tilde{A}_{-}(-q\tau)\, \hat{a}_{-}(\vbf{q}) + \tilde{A}_{-}(-q\tau)^{*}\, \hat{a}^{\dagger}_{-}(-\vbf{q}) \right].
\end{align}
\section{Scalar and tensor perturbations induced by the vector fields}\label{AppB}
In this appendix, we study the sourced power spectra of tensor and scalar perturbations that are generated by the excited vector fields. For the clarity of the discussion, we will first derive the corresponding expressions suitable for the full numerical evaluation using which one can obtain the results derived in the recent literature that we closely follow.

\subsection*{Sourced tensor power spectra}
We start with the sourced contribution to the metric fluctuations which can be written using the Green's function formalism as \cite{Namba:2015gja}:
\beq\label{hs}
\hat{h}^{(\rm s)}_\lambda (\tau, \vbf{k}) = \frac{2}{a(\tau)\Mp} \int \d \tau' \, G_k(\tau,\tau')\, \hat{J}_{\lambda}(\tau',\vbf{k}),
\eeq
where $\hat{J}_\lambda$ is the source term bilinear in the electromagnetic fields and $G_k$ is retarded the Green's function for the vacuum equation of motion of the metric:
\beq\label{Gk}
G_{k}\left(\tau,\tau^{\prime}\right)=\Theta\left(\tau-\tau^{\prime}\right)\frac{\pi}{2}\sqrt{\tau\tau^{\prime}}\left[J_{3/2}(-k\tau)Y_{3/2}\left(-k\tau^{\prime}\right)-Y_{3/2}(-k\tau)J_{3/2}\left(-k\tau^{\prime}\right)\right].
\eeq
Since we are interested in the late time metric perturbation, $-k\tau \ll 1$, the latter can be approximated as \cite{Sorbo:2011rz} 
\beq\label{GkL}
G_k\left(\tau, \tau^{\prime}\right) \simeq \Theta\left(\tau-\tau^{\prime}\right)\sqrt{\frac{\pi}{2}}\frac{\sqrt{\tau}\tau^{\prime}}{(-k\tau)^{3/2}}J_{3/2}(-k\tau^{\prime}) = \frac{\Theta\left(\tau-\tau^{\prime}\right)}{k^3 \tau \tau^{\prime}}\left[k \tau^{\prime} \cos \left(k \tau^{\prime}\right)-\sin \left(k \tau^{\prime}\right)\right].
\eeq
The source term on the other hand is given by as a convolution in the electromagnetic sources in \eqref{EB} \cite{Namba:2015gja, Ozsoy:2020ccy},
\begin{align}
\nn \hat{J}_\lambda (\tau',\vbf{k}) &= -\frac{1}{2\,\Mp\, a(\tau')}  \int \frac{\d^3 \vbf{p}}{(2\pi)^{3/2}}\, \epsilon_\lambda \left[\vbf{k},\vbf{k}-\vbf{p},\vbf{p}\right] \sqrt{|\vbf{k}-\vbf{p}| |\vbf{p}|}\\
&\quad\quad\quad\quad\quad\quad\quad\quad\quad\times \left[\hat{\mathcal{O}}_E(\tau', \vbf{k}-\vbf{p})\,\hat{\mathcal{O}}_E(\tau', \vbf{p}) + \hat{\mathcal{O}}_B(\tau', \vbf{k}-\vbf{p})\,\hat{\mathcal{O}}_B(\tau', \vbf{p})\right],
\end{align}
where we defined a product of polarization vectors that is symmetric in its last two arguments:
\beq
\epsilon_\lambda\left[\vbf{k},\vbf{q},\vbf{q}'\right] = \Pi_{ij,\lambda}(\vbf{k})\, \epsilon^{(-)}_i (\vbf{q}) \,\epsilon^{(-)}_j (\vbf{q}') = \epsilon^{(\lambda)}_i(\vbf{k})^{*}\,\epsilon^{(-)}_i (\vbf{q}) \,\epsilon^{(\lambda)}_j(\vbf{k})^{*}\,\epsilon^{(-)}_j (\vbf{q}').
\eeq
Using Wick's theorem we calculate the contractions involved with $\hat{\mathcal{O}}_E$ and $\hat{\mathcal{O}}_B$ to obtain the two point correlator of the tensor source term as
\begin{align}\label{JLJL}
\nn \langle \hat{J}_\lambda (\tau',\vbf{k}) \hat{J}_\lambda' (\tau'',\vbf{k}') \rangle &= \frac{\delta(\vbf{k}+\vbf{k}')\,\delta_{\lambda\lambda'}}{32 \Mp^2 \,a(\tau')a(\tau'')} \int \frac{\d^3\vbf{p}}{(2\pi)^3} \left(1 - \lambda \hat{\vbf{k}}.\hat{\vbf{p}}\right)^2 \left(1 - \lambda \frac{\hat{\vbf{k}}.(\vbf{k}-\vbf{p})}{|\vbf{k}-\vbf{p}|}\right)^2 |\vbf{k}-\vbf{p}|\, |\vbf{p}|\\
&\quad\quad\quad\quad\quad\quad\quad\quad\quad \times f_A(-|\vbf{k}-\vbf{p}|\tau',-|\vbf{p}|\tau')\, f_A(-|\vbf{k}-\vbf{p}|\tau'',-|\vbf{p}|\tau'')^{*},
\end{align}
where we utilized the identity
\beq
\int \d \phi\,\, \epsilon_\lambda\left[\vbf{k},\vbf{q},\vbf{q}'\right] \epsilon_\lambda\left[\vbf{k},\vbf{q},\vbf{q}'\right]^{*} = \frac{\delta_{\lambda \lambda'}}{16} \int \d \phi \left(1 - \lambda \hat{\vbf{k}}.\hat{\vbf{q}}\right)^2 \, \left(1 - \lambda \hat{\vbf{k}}.\hat{\vbf{q}'}\right)^2 
\eeq
and defined 
\beq\label{dfA}
f_A(-q\tau,-p\tau) \equiv \tilde{A}'_{-}(-q\tau)\, \tilde{A}'_{-}(-p\tau)  + \tilde{A}_{-}(-q\tau)\,\tilde{A}_{-}(-p\tau).
\eeq
Finally noting the late time auto-correlator of the sourced tensor perturbation \eqref{hs},
\beq
\langle \hat{h}^{(\rm s)}_\lambda (\vbf{k})\hat{h}^{(\rm s)}_{\lambda'} (\vbf{k}') \rangle = \frac{4}{a(\tau)^2 \Mp^2} \int \d \tau' G_k(\tau \to 0,\tau')\, \int \d \tau'' \,G_k(\tau\to 0,\tau'')\, \langle \hat{J}_\lambda (\tau',\vbf{k}) \hat{J}_{\lambda'} (\tau'',\vbf{k}') \rangle,
\eeq
we combine our findings to obtain the power spectrum (see \eg \eqref{psdef}) of tensor perturbations as
\begin{align}\label{PL}
\nn 
\mathcal{P}^{(\rm s)}_\lambda (k) &= \frac{H^4}{64\pi^4\Mp^4} \int \d \tilde{p} \int \d \eta \, \left(1 - \lambda \eta \right)^2 \left(1 - \lambda\frac{(1-\tilde{p}\eta)}{\sqrt{1-2\tilde{p}\eta + \tilde{p}^2}} \right)^2 \, \tilde{p}^3 \sqrt{1-2\tilde{p}\eta + \tilde{p}^2}\\
&\quad\quad\quad\quad\quad\quad\quad\quad \times \Bigg|\int_0^{x_{\rm cut}} \d x' \left[x' \cos x' - \sin x'\right] f_A\left(\sqrt{1-2\tilde{p}\eta + \tilde{p}^2} x', \tilde{p}x'\right) \Bigg|^2,
\end{align}
where we wrote the loop measure in \eqref{JLJL} in terms of the cosine angle $\eta$ between $\vbf{k}$ and $\vbf{p}$ using the azimuthal symmetry and defined $\tilde{p} = |\vbf{p}|/k$. This result  can be recast as in \eqref{SC} by re-writing the pre-factor in \eqref{PL} in terms of the vacuum power spectrum of tensor perturbations using \eqref{psv} and defining 
\begin{align}\label{f2Ln}
f_{2,\lambda}\left(\xi_*, x_*, \delta\right)  &= \int \d \tilde{p} \int \d \eta \, \left(1 - \lambda \eta \right)^2 \left(1 - \lambda\frac{(1-\tilde{p}\eta)}{\sqrt{1-2\tilde{p}\eta + \tilde{p}^2}} \right)^2 \, \tilde{p}^3 \sqrt{1-2\tilde{p}\eta + \tilde{p}^2}\\ \nn
&\quad\quad\quad\quad\quad\quad\quad\quad \times \Bigg|\int_0^{x_{\rm cut}} \d x' \left[x' \cos x' - \sin x'\right] f_A\left(\sqrt{1-2\tilde{p}\eta + \tilde{p}^2} x', \tilde{p}x'\right) \Bigg|^2.
\end{align}
We stress here that although the expression above does not appear to have an explicit dependence on the parameters $\xi_*, x_* = k/k_*, \delta$ it does so due to dependence of the effective coupling $\xi$ \eqref{xip} to these parameters as we mentioned in Appendix \ref{AppA}. For the full numerical evaluation of sourced tensor power spectrum we use the expression \eqref{f2Ln} (and \eqref{dfA}) with the numerical evaluation of the vector mode functions we discussed in the previous section. This procedure requires a cut-off procedure that we implement through the upper limit of the time integral $x_{\rm cut}$ in order to eliminate the non-physical vacuum fluctuations of the vector fields deep inside the horizon. 

\medskip
\noindent{\bf Sourced tensor spectrum in the WKB approximation.} If we utilize the WKB approximated solutions \eqref{Amsol}, the tensor spectrum produced by the gauge field fluctuations reduces to the results derived in the literature \cite{Namba:2015gja,Ozsoy:2020ccy} as we illustrate now. Using the real solution in \eqref{Amsol} with \eqref{f2Ln} and \eqref{rd}, one can derive the function $f_A$ \eqref{dfA} that appear in \eqref{f2Ln} in the WKB approximation as 
\begin{align}\label{fAWKB}
f_A(-\tilde{q} x',-\tilde{p}x') &= \frac{\sqrt{2}\, N(\xi_*, \tilde{q}x_*, \delta)\,  N(\xi_*, \tilde{p} x_*, \delta)}{(\tilde{q}\,\tilde{p})^{1/4}} \left(\sqrt{\frac{\xi(x')}{x'}} + \sqrt{\frac{x'}{\xi(x')}}\frac{\sqrt{\tilde{q}\, \tilde{p}}}{2}  \right)\\ \nn
&\quad\quad\quad\quad\quad\quad\quad\quad\quad\quad\quad\quad\quad\quad \times \exp\left[-2E(x')\sqrt{2\xi_* x_*}\left(\sqrt{\tilde{q}}+\sqrt{\tilde{p}}\right)\right].
\end{align}
Plugging \eqref{fAWKB} in \eqref{f2Ln} we get  
\begin{align}\label{f2Lwkb}
f_{2,\lambda}(\xi_{*},x_{*},\delta) &= 2\int \mathrm{d}{\tilde{p}}\int\mathrm{d}\eta\, \left(1 - \lambda \eta \right)^2 \left(1 - \lambda\frac{(1-\tilde{p}\eta)}{\sqrt{1-2\tilde{p}\eta + \tilde{p^2}}} \right)^2 {\tilde{p}}^{5/2} \left(1-2{\tilde{p}}\eta+{\tilde{p}}^{2}\right)^{1/4}\\ \nn
&\quad\quad \times \,N^{2}\left(\xi_{*},\sqrt{1-2\tilde{p}\eta+\tilde{p}^{2}}x_{*},\delta\right)N^{2}\left(\xi_{*},\tilde{p}x_{*},\delta\right)\mathcal{I}^{2}\left[\xi_{*},x_{*},\delta,\tilde{p},\sqrt{1-2\tilde{p}\eta+\tilde{p}^{2}}\right],
\end{align}
where we defined the time integral of the sources as
\begin{align}\label{IT}
\mathcal{I}\bigg[\xi_{*},x_{*},\delta,\tilde{p},\tilde{q}\bigg]&\equiv
\int_{0}^{x_{\rm cut}}\mathrm{d}x^{\prime}\left(x^{\prime}\cos
x^{\prime}-\sin x^{\prime}\right)\left[\sqrt{\frac{\xi\left(x^{\prime}\right)}{x^{\prime}}} + \frac{\sqrt{\tilde{q}\,\tilde{p}}}{2}\sqrt{\frac{x^{\prime}}{\xi\left(x^{\prime}\right)}}\right]\\ \nn
&\quad\quad\quad\quad\quad\quad\quad\quad\quad\quad\quad\quad\quad\quad \quad\quad \times \exp\left[-2E(x')\sqrt{2\xi_* x_*}(\sqrt{\tilde{q}} + \sqrt{\tilde{p}})\right],
\end{align}
with $x_{\rm cut}$ introduced to exclude non-physical vacuum modes. For the M1 model, the exclusion of the non-physical vacuum modes inside the integral \eqref{IT} above is automatically implemented when we adopt the approximate WKB solution (which is the growing solution outside the horizon) in \eqref{Amsol} which is decaying deep inside the horizon $x \to \infty$ noting the functional form of $E(x)$ given in the first line of \eqref{Eform}. On the other hand, the same situation does not hold for the M2 model as $E(x)\to \infty$ for $x \to \infty$ leading to divergent time integral of the sources in \eqref{IT}. To regularize the integral \eqref{IT}, we restrict the upper limit of the time integration domain to $x_{\rm cut} = \xi_*$ which (roughly) corresponds to the time when all the modes of interest is at the onset of tachyonic instability \cite{Peloso:2016gqs}. Since the arguments of the gauge mode functions are re-scaled with factors of $\tilde{q}$ and $|\hat{k}-\vec{\tilde{q}}| = \sqrt{1-2\eta \tilde{q} + \tilde{q}^2}$ in the loop integral \eqref{f2Lwkb}, a more proper choice of the cut-off is given by 
\beq
x_{\rm cut} = \xi_* \times {\rm Min}\left[1, \frac{1}{\tilde{q}}, \frac{1}{|\hat{k}-\vec{\tilde{q}}|}\right]
\eeq
to accurately isolate the divergent part. This is the cut-off we implement in this work when necessary including the fully numerical approach we undertake. 
For the evaluation of the sourced tensor power spectrum in the WKB approximation, we use the expression \eqref{f2Lwkb} with the numerically obtained normalization factors $N(\xi_*, x_*, \delta)$ implementing the cut-off procedure we describe above. As we emphasize in this work, this procedure leads to significantly different results for the scale dependence of the sourced tensor power spectrum as compared to the common practice which assumes a log-normal fitting form \eqref{Nform} for $N(\xi_*, x_*, \delta)$.

\medskip

{{\noindent{\bf Dominant contribution to the integral and UV cut-off insensitivity.}\\ We devote this sub-section to dive deeper into some important features of the  above-mentioned correlators. For illustrative purposes, we will use Fig.~\ref{fig:xi-evolution} as a typical profile of the particle production parameter $\xi$ considered in this work. Let us denote by $k_A$ the momentum of an internal gauge field perturbation which leaves the horizon at the moment $N_A$ and likewise for other time points shown in Fig.~\ref{fig:xi-evolution}: $k_*$ at $N_*$ and $k_B$ at $N_B$, where $N_A < N_* < N_B$. Next, let us consider an external gravitational wave of momentum $k_A$ which, following our definitions above, leaves the horizon at moment $N_A$. In principle, such a GW can be sourced by gauge fields of all possible (internal) momenta $p$, from very small (those that left the horizon much earlier than $N_A$)  to very large momenta (those that leave the horizon after $N_B$). The conventional wisdom so far has been that the dominant contribution to a sourced GW of external momentum $k_A$ comes from gauge fields of internal momenta in the neighborhood of $p\sim k_A$ since much larger or smaller momenta will suffer from significant phase-space suppression.

If that were the case in the present context, one could rightfully point out that our cut-off procedure would not be reliable in light of recent works such as \cite{Durrer:2024ibi}. Therein, it has been conclusively shown that, for small particle production parameters (as is the case at $N_A$), the simple cut-off procedure is ambiguous and one ought to use a more robust regularization and renormalization prescription to extract the final result as done in \cite{Ballardini:2019rqh,Animali:2022lig}. One of the most important results of this work is that such conventional wisdom does in fact not apply when $\xi$ displays a bumpy profile as in Fig.~\ref{fig:xi-evolution}. Instead, in such cases all sourced GWs, regardless of their momenta (and correspondingly horizon-crossing time), and by extension the value of the particle production parameter $\xi$ in that moment, are universally sourced by gauge fields of momentum in the neighborhood of $k_*$ which crossed the horizon at around moment $N_*$. It follows that all GWs are correspondingly enhanced by the maximum $\xi$ value, i.e. $\xi_*$ , which is sufficiently large to make  our cut-off procedure reliable. 

\begin{figure}[t!]
\begin{center}
\includegraphics[scale=0.6]{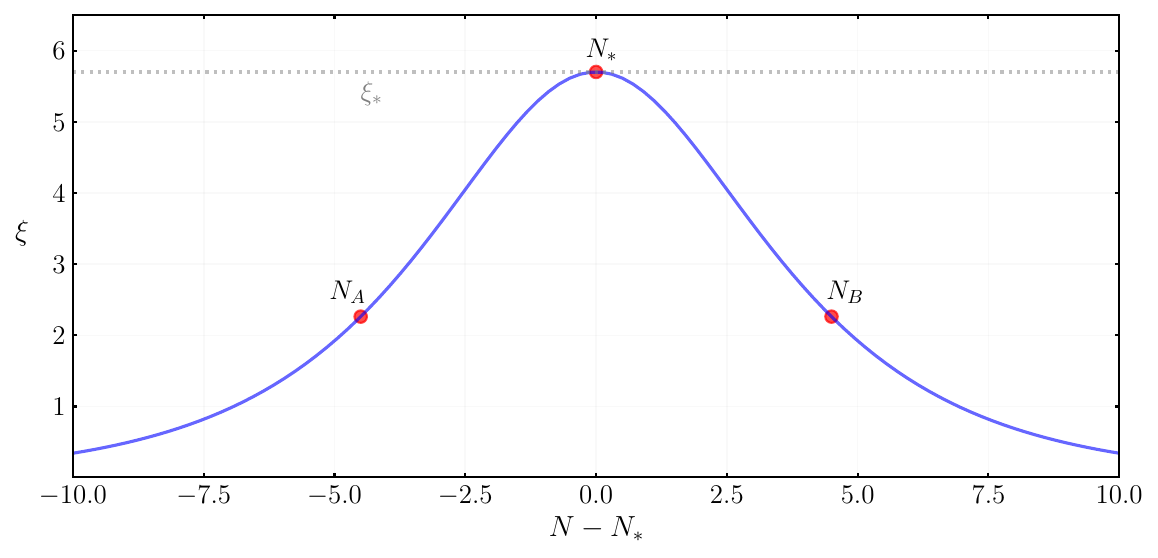}
\end{center}
\vspace*{-5mm}\caption{{The typical time-evolution of the ``bumpy'' particle production parameter  in the cases  considered in this work. The labels $N_A$, $N_*$ and $N_B$ denote three distinct moments in time: one for which $\xi$ has not yet reached its maximum value, one for which $\xi$ is exactly at the maximum $\xi_*$, and one for which $\xi_*$ is decreasing from the maximum. The horizontal axis denotes the number of e-folds relative to $N_*\,$.}\label{fig:xi-evolution}}
\end{figure}

One can see that this is the case by considering that the sourced GWs  are ultimately integrals over four powers of the gauge mode functions. Parametrically then, the amplitude of the sourced GW is proportional to $P_h \propto {\rm PSS}\times  {\rm e}^{4\pi \xi}$, where PSS stands for Phase Space Suppression. Since gauge fields of all possible momenta source the GWs, there is a trade-off between the size of the particle production parameter and the corresponding phase space suppression due to the hierarchy between the external (GW) and internal (gauge field) momentum. Depending on the width and the height of the peak in the $\xi$ profile, it is possible to identify an example for which the phase space suppression is the subdominant effect. It is instead the enormous occupation number attained by the gauge field at moment $N_*$ that dominates the production of GWs of all scales. Pointing out that such cases exists and they respect the adiabatic-evolution conditions for $\xi$ is one of the main points of the present work.

 As a side note, we find it important to stress that the GWs of (e.g.) momentum $k_A$ are still sourced by gauge fields of momentum $p \sim k_A$, for which there is an ambiguity regrading the regularization-renormalization procedure. However, any such contribution coming from small gauge field momenta is a negligible part of the total result. We therefore  do not expect our results to significantly change upon implementing a more sophisticated approach than the simple cut-off we employ here. 

As a final note, and also a segue to the next sub-section, it is worthwhile to note that (i) GWs of momentum $k_A$ are predominantly sourced by gauge field momenta that are much greater than $k_A$, (ii) GWs of momentum $k_*$ are predominantly sourced by gauge fields in the neighborhood of $k_*$, and finally (iii) GWs of momentum $k_B$ are predominantly sourced by at least one gauge field whose momentum is far smaller than $k_B$. We will see that this hierarchy between the size of the external momentum $k$ and the loop momenta $p$ (as well as their angles) is precisely the physical origin of the scale-dependent chirality which we highlight in this work.}
\medskip

{\noindent{\bf Revisiting the origin of scale-dependent GW chirality.} We find it useful to provide here a rather condensed discussion on the emergence of scale-dependent parity violation of the produced GW signal. More details have been provided in section \ref{sec4p2}. 
For this purpose, we turn our attention to the part of the two-point GW correlator called $f_{2,\lambda}$ that exhibits scale-dependence. The variable  $x_* = k/k_*$, in turn, parametrizes the size of the external tensor modes in eq. \eqref{f2Lwkb}. In this expression we can identify three important ingredients that can help us gain a better understanding of the origin of scale-dependent GW chirality: 
\begin{enumerate}
\item The dominant support to the tensor correlator \eqref{f2Lwkb} comes from the peak of the scale-dependent vector normalization factors $N$ (at fixed $\xi_*$ and $\delta$), which depend on the normalized loop momenta $\tilde p$, the cosine angle $\eta$ and the wave-number $x_* = k/k_*$ of the external GW.
\item The part of the tensor correlator that determines the amplitude of the different GW polarization states is given by 
\beq
p_\lambda(\tilde p, \eta) \equiv \left(1 - \lambda \eta \right)^2 \left(1 - \lambda\frac{(1-\tilde{p}\eta)}{\sqrt{1-2\tilde{p}\eta + \tilde{p^2}}} \right)^2 \; .
\label{eq:polarization-factor}
\eeq
\item A tensor mode with momentum $\vec{k}$ is produced via a loop of two vector modes with momenta $\vec{k} - \vec{p}$ and $\vec{p}$,  such that the momentum is conserved at each vertex.  
\end{enumerate}
We infer from the first point above that to produce a soft GW mode with $x_* = k/k_* \ll 1$, the vector modes running in the loop must satisfy $\tilde{p} \gg 1$. This becomes apparent from the expression \eqref{f2Lwkb} by noticing the fact the integral gets most of its support from the peak of the vector normalization factor $N$ where $\xi_*$ is at its maximum (see Figure \ref{fig:norms}).  In this limit ($\tilde{p} \to \infty$),
momentum conservation (see point 3 above) at the vertex singles out the anti-aligned configuration\footnote{Notice that such a configuration can be achieved for any angle $\eta$ since $\eta$ is the angle between the external $\vec{k}$ and internal momenta $\vec{p}$.} between the two internal gauge modes. The key point is that this configuration is not sensitive to left/right handed GW modes as  each  carries equal weight (see point 2 above):
\beq
\lim_{\tilde{p} \to \infty} \frac{p_-(\tilde p, \eta)}{p_+(\tilde p, \eta)} \quad \to  \quad 1 + \frac{4}{\tilde{p}}\quad \to  \quad 1\;\;.
\eeq
In other words, in the anti-aligned configuration the ratio of the phase space available for the two GW polarization states  approaches unity, $p_+ / p_{-} \to 1$, leading to a suppressed chirality in the IR, $x_* = k/k_* \ll 1$.

Similarly, the GW spectrum \eqref{f2Lwkb} in the UV ($x_* = k/k_* \gg 1$) only has  support for one of the two loop momenta satisfying $\tilde{p} \ll 1$ as the vector sources are maximal only for $\tilde{p}\,x_* \sim 1$. In this limit ($\tilde{p} \to 0$), momentum conservation allows for any angle between the internal gauge modes. However, the co-linear configuration allows only one polarization,  forbidding altogether the production of the opposite one. Taking the limit again we find
\beq
\lim_{\tilde{p} \to 0} \frac{p_-(\tilde p, \eta)}{p_+(\tilde p, \eta)} \quad \to  \quad \frac{16}{(\eta-1)^4\tilde{p}^4} \quad \to  \quad \infty\;\;,
\eeq
which  confirms our findings: GW spectra exhibit maximal chirality for scales following the peak, \ie scales such that $k/k_* \gg 1$ (see Figs. \ref{fig:dchiM1} and \ref{fig:dchiM2}). }

\begin{figure}[t!]
\begin{center}
\includegraphics[scale=0.75]{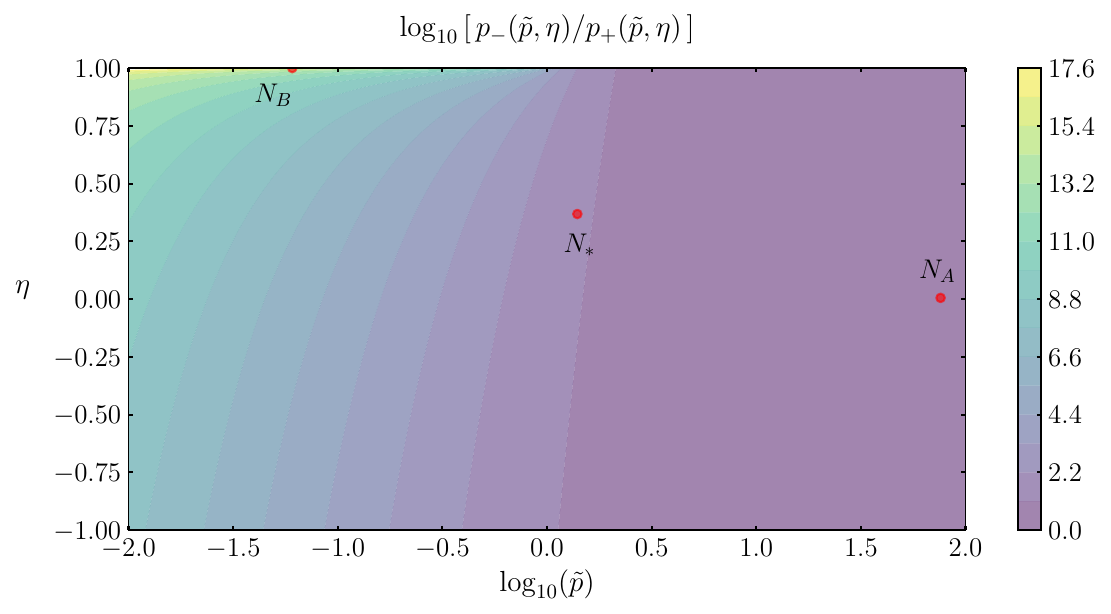}
\end{center}
\vspace*{-5mm}\caption{{A plot of the ratio of the two polarization-dependent factors defined in (\ref{eq:polarization-factor}). The greater the hierarchy between internal and external momentum ($\tilde{p}\gg 1$), the more the sourcing of the GWs becomes chirality-insensitive. The red dots mark the points in the phase space that dominate the sourcing of GWs that left the horizon in the three moments marked in Fig.~\ref{fig:xi-evolution}.}\label{fig:Contour-Plot}}
\end{figure}

In order to emphasise our results, we plot the ratio of the contributions from the two different polarizations in Fig.~\ref{fig:Contour-Plot}, a representation fully consistent with the limits evaluated analytically above. The final GW power will depend on the product between the polarization dependent factor defined in (\ref{eq:polarization-factor}) and the polarization independent residual factor in the integrand of eq.~(\ref{f2Ln}). This residual factor is precisely what controls which part of the phase space $(\tilde{p},\eta)$ will dominate the GW sourcing. We identified the point in phase space at which the production peaks for the three sample moments in time in the toy model of Fig.~\ref{fig:xi-evolution}. We indicate these points with red dots in Fig.~\ref{fig:Contour-Plot} to show more explicitly how the hierarchy between the external and internal momenta have a direct impact on the  GW chirality. Indeed, based on the locations of peak production, one would expect the chirality of the sourced GWs to increase as one moves from $N_A$ to $N_*$ and finally to $N_B$. Such a behaviour is reflected in the results we report in the main text.}

\subsection*{Sourced scalar power spectra}

We continue with the contribution to the curvature perturbation by the excited vector field fluctuations which can be written in the Green's function formalism as \cite{Namba:2015gja,Ozsoy:2020ccy},
\beq\label{SZ}
\hat{\zeta}^{(s)}(\tau,\vbf{k})\simeq\frac{3\sqrt{2}H\tau}{M_{\mathrm{pl}}}\int
\d\tau^{\prime}\,G_{k}\left(\tau,\tau^{\prime}\right)\frac{\sqrt{\epsilon_{\chi}\left(\tau^{\prime}\right)}}{\tau^{\prime2}}\int
\d\tau^{\prime\prime}\,G_{k}\left(\tau^{\prime},\tau^{\prime\prime}\right)\hat{J}_{\chi}\left(\tau^{\prime\prime},\vbf{k}\right),
\eeq
where $\epsilon_\chi = \dot{\chi}^2 / (2 H^2 \Mp^2)$ is the slow-roll parameter of the spectator axion and the Green's function takes the same form in \eqref{Gk} as the canonical axion and inflaton perturbations (see \eg \eqref{usgm} and \eqref{uphi}) obey the same homogeneous equation. The source $\hat{J}_\chi$ in \eqref{SZ} is given by \cite{Namba:2015gja}
\begin{align}\label{Jx}
\nn \hat{J}_{\chi}(\tau,\vbf{k})&=\frac{\lambda\, a(\tau)^{3}}{f}\int\frac{\mathrm{d}^{3}\vbf{p}}{(2\pi)^{3/2}}\;\hat{E}_{i}(\tau,\vbf{k}-\vbf{p})\;\hat{B}_{i}(\tau,\vbf{p}),\\
&= - \fr{\lambda}{2f\, a(\tau)} \int \frac{\mathrm{d}^{3}\vbf{p}}{(2\pi)^{3/2}} \,\epsilon^{(-)}_i (\vbf{k}-\vbf{p})\, \epsilon^{(-)}_i (\vbf{p})\, \hat{\mathcal{O}}_E(\tau, \vbf{k}-\vbf{p})\,\hat{\mathcal{O}}_B(\tau, \vbf{p}),
\end{align}
where the operators $\hat{\mathcal{O}}_{E/B}$ are defined as in \eqref{OEB}. Using Wick’s theorem, it is then straightforward to compute the contractions of $\hat{\mathcal{O}}_E$ and $\hat{\mathcal{O}}_B$, to obtain two point correlator of the scalar source \eqref{Jx} as
\begin{align}\label{JxJx}
\langle \hat{J}_\chi (\tau,\vbf{k}) \hat{J}_\chi (\tau',\vbf{k}') \rangle &= \frac{\delta(\vbf{k}+\vbf{k}')\,\lambda^2}{8 f^2 \,a(\tau)a(\tau')} \int \frac{\d^3\vbf{p}}{(2\pi)^3} \left(1 - \frac{\hat{\vbf{p}}.(\vbf{k}-\vbf{p})}{|\vbf{k}-\vbf{p}|}\right)^2 |\vbf{k}-\vbf{p}|\, |\vbf{p}|\\ \nn
&\quad\quad\quad\quad\quad\quad\quad\quad\quad \times \tilde{A}'_{-}(-|\vbf{k}-\vbf{p}|\tau)\, \tilde{A}'_{-}(-|\vbf{k}-\vbf{p}|\tau')\,\tilde{A}_{-}(-|\vbf{p}|\tau)\,\tilde{A}_{-}(-|\vbf{p}|\tau'),
\end{align}
where we utilized the following identity $\left|\epsilon^{(-)}_i(\vbf{q})\epsilon^{(-)}_i(\vbf{q}')\right|^2  = \left(1 - \hat{\vbf{q}}.\hat{\vbf{q}}' \right)^2/4$. 
Using \eqref{JxJx}, we then take the two point function of the sourced curvature perturbation \eqref{SZ} following the same steps we carried for the tensor perturbations. In this way, the sourced power spectrum of curvature fluctuations takes the following form 
\begin{align}
\mathcal{P}^{(\rm s)}_{\zeta}(k) = \left[\epsilon_\phi \mathcal{P}^{(\rm v)}_{\zeta}(k)\right]^2 f_{2,\zeta} (\xi_*, x_*, \delta),
\end{align}
where the the part that characterizes the influence of scale dependent particle production in the gauge sector can be written as
\begin{align}\label{f2z}
f_{2,\zeta} (\xi_*, x_*, \delta) &= \frac{9\pi^3 \xi_*^2}{2} \int \d \tilde{p} \int \d \eta \left(1 - \frac{(\eta - \tilde{p})}{\sqrt{1-2\tilde{p}\eta + \tilde{p^2}}}\right)^2 \tilde{p}^3 \sqrt{1-2\tilde{p}\eta + \tilde{p}^2}\\ \nn
&\quad\quad\quad \times \bigg| \int \frac{\d x'}{x'} J_{3/2}(x')\sqrt{\frac{\epsilon_{\chi}(x')}{\epsilon_{\chi,*}}} \int \d x'' {x''}^{3/2}\left[J_{3/2}(x')Y_{3/2}(x'') - Y_{3/2}(x')J_{3/2}(x'')\right]\\ \nn
& \quad\quad\quad\quad\quad\quad\quad\quad\quad\quad\quad\quad\quad\quad\quad\quad\quad\quad\quad\quad \times \tilde{A}'_{-}\left(\sqrt{1-2\tilde{p}\eta + \tilde{p}^2}x''\right)\, \tilde{A}_{-}\left(\tilde{p} x''\right) \bigg|^2.
\end{align}
Note that here, we only take the late time limit of the outer most Green's function in \eqref{SZ} using \eqref{GkL} while keeping the full expression \eqref{Gk} for the inner Green's function.
Following the discussion we provide in Appendix \ref{AppA}, to fully numerically compute \eqref{f2z}, we also require 
\beq
\xi^2 \equiv \frac{\lambda^2 \Mp^2}{2 f^2} \epsilon_\chi \quad \longrightarrow \quad \sqrt{\frac{\epsilon_\chi}{\epsilon_{\chi,*}}} = \frac{\xi}{\xi_*},
\eeq
where $\xi$ is given as in \eqref{xip}.

\medskip
\noindent{\bf Sourced curvature spectrum in the WKB approximation.} For completeness, below we provide the expression for the sourced scalar power spectrum in the WKB approximation. For this purpose, we just need to re-write the last line of \eqref{f2z} using \eqref{ir} and \eqref{Amsol} as
\begin{align}
\tilde{A}'_{-}\left(\tilde{q} x''\right)\, \tilde{A}_{-}\left(\tilde{p} x''\right) = \frac{\tilde{p}^{1/4}}{\tilde{q}^{1/4}}\, N(\xi_*, \tilde{q}x_*, \delta)\,  N(\xi_*, \tilde{p} x_*, \delta)\,\exp\left[-2E(x'')\sqrt{2\xi_* x_*}\left(\sqrt{\tilde{q}}+\sqrt{\tilde{p}}\right)\right],
\end{align}
for arbitrary arguments of the mode functions.
Inserting this expression in \eqref{f2z}, we obtain the corresponding result in the WKB approximation as 
\begin{align}\label{f2zwkb}
f_{2,\zeta} (\xi_*, x_*, \delta) &= \frac{9\pi^3 \xi_*^2}{2} \int \d \tilde{p} \int \d \eta \left(1 - \frac{(\eta - \tilde{p})}{\sqrt{1-2\tilde{p}\eta + \tilde{p^2}}}\right)^2 \tilde{p}^{7/2} \left(1-2\tilde{p}\eta + \tilde{p}^2\right)^{1/4}\\ \nn
&\quad \times N(\xi_*, \sqrt{1-2\tilde{p}\eta + \tilde{p}^2}x_*, \delta)^2\,  N(\xi_*, \tilde{p} x_*, \delta)^2\,\mathcal{I}_\zeta \left[\xi_*, x_*, \delta, (1-2\tilde{p}\eta + \tilde{p}^2)^{1/4} + \sqrt{\tilde{p}} \right]^2,
\end{align}
where we defined the time integral of the vector mode functions as 
\begin{align}\label{Izwkb}
\nn \mathcal{I}_\zeta \left[\xi_*, x_*, \delta, Q \right] &\equiv 
 \int \frac{\d x'}{x'} J_{3/2}(x')\sqrt{\frac{\epsilon_{\chi}(x')}{\epsilon_{\chi,*}}} \int \d x'' {x''}^{3/2}\left[J_{3/2}(x')Y_{3/2}(x'') - Y_{3/2}(x')J_{3/2}(x'')\right]\\ 
 & \quad\quad\quad\quad\quad\quad\quad\quad\quad\quad\quad\quad\quad\quad\quad\quad\quad \times \exp\left[-2E(x'')\sqrt{2\xi_* x_*}Q\right].
\end{align}
To evaluate the sourced power spectrum within WKB method, we use \eqref{Izwkb} and \eqref{f2zwkb} with the numerically obtained normalization factors $N(\xi_*, x_{*}, \delta)$ we describe in the previous section to discuss phenomenological implications of the resulting signal (see section \ref{sec4}).

\end{appendix}
\newpage
\addcontentsline{toc}{section}{References}
\bibliographystyle{utphys}
\bibliography{paper2}
\end{document}